\documentclass[aps,pra,twocolumn,amsmath,amssymb,superscriptaddress,floatfix]{revtex4-1}

\usepackage{graphicx}
\usepackage{multirow}
\usepackage{color}
\usepackage{bm}

\renewcommand{\Vec}[1]{\bm{#1}}

\def\t#1{\textrm{#1}}

\def\n{\nonumber \\ }
\def\tensor{\otimes}

\definecolor{purple}{rgb}{0.7, 0.0, 0.7}

\begin{document}

\title{
Floquet topological phases
protected by time glide symmetry
}

\author{Takahiro Morimoto}
\affiliation{Department of Physics,
University of California, Berkeley, CA 94720}

\author{Hoi Chun Po}
\affiliation{Department of Physics, University of California, Berkeley, CA 94720}
\affiliation{Department of Physics, Harvard University, Cambridge MA 02138}

\author{Ashvin Vishwanath}
\affiliation{Department of Physics, University of California, Berkeley, CA 94720}
\affiliation{Department of Physics, Harvard University, Cambridge MA 02138}

\date{\today}

\begin{abstract}
We study Floquet topological phases in periodically driven systems that are protected by ``time glide symmetry'', a combination of reflection and half time period translation. Time glide symmetry is an analog of glide symmetry with partial time translation replacing the partial space translation, and hence, is an intrinsically dynamical symmetry which may be engineered in periodically driven systems by exploiting the controllability of driving. We present lattice models of time glide symmetric Floquet topological insulators in two and three dimensions. The topological numbers characterizing those Floquet topological phases are derived from the half period time evolution operator along with time glide operator. Moreover, we classify Floquet topological phases protected by time glide symmetry in general dimensions using a Clifford algebra approach. 
The obtained classification table is similar to that for topological crystalline insulators protected by static reflection symmetry, but shows nontrivial entries in different combination of symmetries, which clarifies that time glide symmetric Floquet topological phases are a distinct set of topological phases from topological crystalline insulators.  We also classify Floquet topological phases with ``time screw symmetry,'' defined as a two-fold spatial rotation accompanied by half-period time translation. 
\end{abstract}

\maketitle

\section{Introduction}
The discovery of topological insulators has revived the prominence of topology in the frontier studies of electronic systems \cite{hasan-kane10,qi-zhang-rmp11}.
A major driving force behind this new wave of development is the realization that symmetries can lead to new topologies. For instance, although a two-dimensional time-reversal-symmetric band insulator necessarily carries a vanishing Chern number, they can still carry a nontrivial $\mathbb Z_2$-valued topological invariant and exhibit the quantum spin Hall effect. As these novel topological distinctions are only well-defined in the presence of certain symmetries, they are generally referred to as `symmetry-protected topological phases' (SPTs), and are now known to exist for both fermionic and bosonic systems with a variety of different symmetries \cite{Chen11,Lu12,XieGuLiuWen}.

Recently, it has been realized that such topological ideas can also be applied to the study of strongly out-of-equilibrium dynamics. In particular, much focus has been placed on Floquet systems -- quantum systems coupled to time-periodic classical drives -- as their phases, defined as classes of long-time behavior, can be systematically studied through the single-period system evolution \cite{Oka,Demler,RudnerLevin,Lindner}. 
General classification results for Floquet systems of non-interacting fermions with on-site symmetry groups have been obtained in Ref.~\cite{Roy16}. They are found to be generally classified in the same way as the corresponding equilibrium systems in the same symmetry class (so called tenfold-way classification \cite{Schnyder08,Ryu10,Kitaev09}), but with a generalization that takes into account the absence of the notion of ground states in a Floquet system.

At a first glance, such similarity between Floquet and static problems may not be unexpected, as in the presence of a Floquet band gap one can define an effective Floquet Hamiltonian and classify it using equilibrium techniques. Yet, such interpretation does not accurately describe the obtained Floquet classification, as intrinsically dynamical phases, which showcase robust topological properties despite a topologically trivial Floquet Hamiltonian, have been discovered. This is exemplified by the `anomalous Floquet Anderson insulators', which are (2+1)D systems that, despite a vanishing bulk Chern numbers, host protected chiral edge modes \cite{RudnerLevin,Lindner}. More recently, it has also been realized that analogous chiral Floquet phases exist for bosonic spin systems, and their classification is distinct from any previously-known equilibrium phases \cite{CFMBL}. 
Furthermore, periodically driven systems with strong interactions constrained by symmetries were shown to host a new class of ``Floquet'' SPTs. These Floquet SPTs are characterized by pumping of equilibrium SPT phases to the surface in each cycle, which were studied in one dimension \cite{Keyserlingk,Keyserlingk2,ElseNayak,PotterMorimotoVishwanath} and also in higher dimensions \cite{PotterMorimoto-2D,RoyHarper-higherD}.  These discoveries demonstrate that Floquet systems are capable of hosting novel topological phases with no equilibrium counterparts.

So far, studies on topological Floquet phases have focused on how conventional symmetry classes, which have played a key role in the classification of equilibrium phases, can lead to novel topological dynamics.
A natural next step forward is to turn our focus from topology to symmetry  -- what are the symmetries that are unique to Floquet systems, and what are the new phases, if any, that they lead to? 

The goal of this work is to initiate the analysis of this problem. We will focus exclusively on `dynamical symmetries' which are unique to Floquet systems \cite{Alon98,Ceccherini,Fregoso13}.
Such symmetries are defined using the discrete time-translation invariance Floquet systems, and can be understood as the space-time analogs of nonsymmorphic spatial symmetries.
For instance, we say the system possesses a `time-glide' symmetry if the instantaneous Hamiltonian $\hat H(t)$ satisfies $ \hat R \hat H(t) \hat R^{-1} = \hat H(t+T/2)$, where $\hat R$ is a spatial reflection and $T$ denotes the Floquet period.
It is compatible with a Floquet system as acting it twice gives $\hat H(t) = \hat H(t+T)$, the defining relation of Floquet problems.
Similarly, one can define the `time-screw' symmetry via $\hat C_n \hat H(t) \hat C_n^{-1} = \hat H(t+T/n)$, where $\hat C_n$ denotes the $n$-fold spatial rotation.
Generally, a Floquet system will be symmetric under a space-time symmetry group featuring nontrivial combination of spatial and temporal operations, similar to how crystals are classified by space groups.

Here, we will take a first step towards understanding the topological consequences of dynamical symmetries in Floquet systems. Specifically, we will focus on free-fermion problems symmetric under either a time-glide or a time-screw squaring to a discrete time translation. 
Such systems represent the simplest setting which demonstrates the existence of new phases arising from dynamical symmetries. We find that their topological characterization cannot be readily interpreted as analogs of equilibrium phases. In particular, Floquet topological phases with time-glide is shown to be distinct from conventional topological crystalline insulators protected by static reflection symmetry \cite{Chiu13,Morimoto13,Shiozaki14}.

This paper is organized as follows: In Sec.~\ref{sec: models}, we will first develop intuition for the definition and consequences of dynamical symmetries by studying explicit (2+1) and (3+1)D lattice models, where we also present explicit derivation of the topological invariants characterizing their phases. In Sec.~\ref{sec: classification time glide}, we will extend the discussions to ($d$+1)D dimensions, and obtain the general classification results analogous to the ten-fold way classification. We will conclude in Sec.~\ref{sec: discussions} by discussing various directions for future works.

\section{Time-glide symmetric Floquet topological phases \label{sec: models}}
We study noninteracting periodically-driven systems with ``time glide symmetry''. The time glide symmetry is an intrinsically dynamical symmetry which is a combination of a reflection symmetry and half time period translation, and is written as
\begin{align}
M_T H(k,t) M_T^\dagger &=H \left(R_i(k),t+\frac T 2 \right).
\end{align}
Here, $R_i$ denotes a reflection along the $i$th direction.
Due to the dynamical nature of the time-glide symmetry, the presence or absence of time glide symmetry can be controlled by designing suitable drivings. In the following, we show examples of Floquet topological phases protected by the time-glide symmetry.

\subsection{2D model of Floquet topological phase with chiral symmetry and time glide symmetry \label{subsec:2d AIII}}

\begin{figure}
\begin{center}
\includegraphics[width=\linewidth]{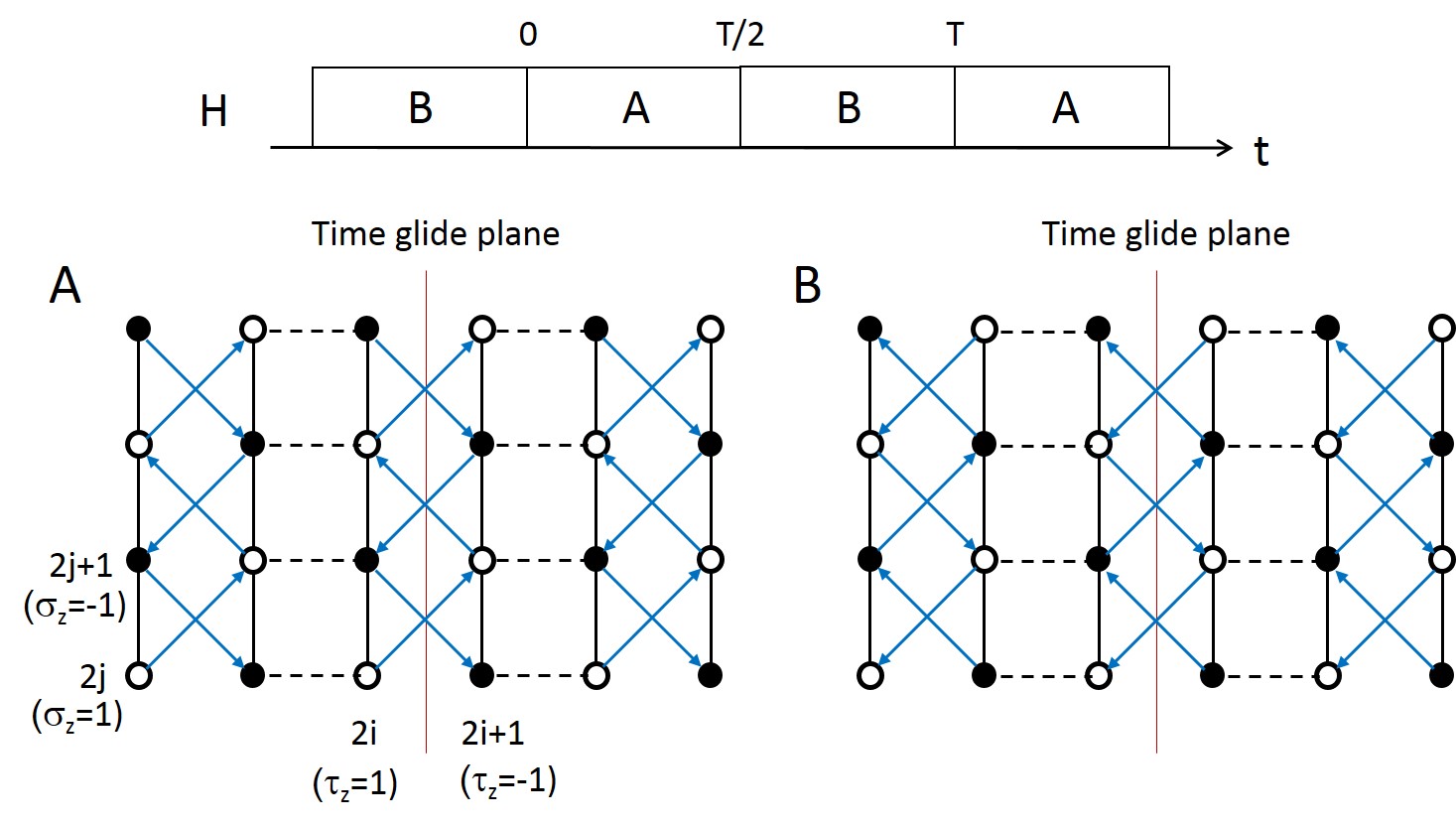}
\end{center}
\caption{Schematic picture of the 2D model with chiral symmetry and time-glide symmetry.
The model consists of a two step drive, denoted by A and B.
Solid and dotted lines represent static hoppings with $t$ and $t'$, respectively.
Blue arrows represent alternating hoppings with the amplitude $it$ ($-it$) along (opposite to) the direction of the arrow.
}
\label{fig: 2D AIII+Tglide}
\end{figure}

First we consider a two-dimensional periodically driven system with chiral symmetry and time-glide symmetry defined on a stack of 1D chains as follows.
The Hamiltonian is given by
\begin{align}
H(t)&=H_\t{intra} + H_\t{inter},
\label{eq: 2d model}
\end{align}
where $H_\t{intra}$ and $H_\t{inter}$ denote intrachain and interchain couplings, respectively,
and are given by
\begin{align}
H_\t{intra}&=
t \sum_{i,j} c_{i,j+1}^\dagger c_{i,j} + \t{h.c.},
\end{align}
and
\begin{align}
H_\t{inter}(t)&= 
t' \sum_{i,j} c_{2i+2,j}^\dagger c_{2i+1,j} \n
&\quad 
+i t \eta(t) \sum_{i,j} (c_{2i+1,2j+1}^\dagger c_{2i,2j} + c_{2i+1,2j}^\dagger c_{2i,2j+1}
\n
&\quad
+c_{2i,2j+1}^\dagger c_{2i+1,2j+2} + c_{2i,2j+2}^\dagger c_{2i+1,2j+1}) 
+ \t{h.c.},
\end{align}
with 
\begin{align}
\eta(t)&=
\begin{cases}
+1, & (0\le t < \frac T 2) \\
-1. & (\frac T 2 \le t < T) 
\end{cases}
\end{align}
Here, $c_{i,j}$ denotes the annihilation operator of an electron at the $j$th site in the $i$th chain. 
In the momentum space representation, this Hamiltonian reads
\begin{align}
H(\bm k, t)&=
2t \sigma_x \cos \frac{k_y}{2} + 2t \eta(t) \sigma_y \tau_y \sin \frac{k_y}{2}
\n
&\quad + t' \tau_x \cos k_x + t' \tau_y \sin k_x,
\end{align}
where Pauli matrices $\sigma_i, \tau_i$ act on the sublattice within the chain and two chains in the unit cell, respectively.
Note that the lattice constant along the $y$-direction is $2$.
This two-step drive is schematically illustrated in Fig.~\ref{fig: 2D AIII+Tglide}. 

The above Hamiltonian $H(t)$ preserves the chiral symmetry as
\begin{align}
\Gamma H(t) \Gamma &= -H(-t), & \Gamma&=\sigma_z \tau_z, 
\label{eq: chiral symmetry}
\end{align}
where sites with $\Gamma=+1$ and $\Gamma=-1$ are illustrated with white and black dots in Fig.~\ref{fig: 2D AIII+Tglide}. 
In particular, the driving at a time slice does not satisfy the chiral symmetry because $\Gamma H_{A/B}\Gamma \not = - H_{A/B}$, where we write $H_A=H(t)$ for $0\le t < T/2$ and $H_B=H(t)$ for $T/2\le t < T$.
Nonetheless, it satisfies the chiral symmetry as a whole function of $t$ as $\Gamma H_A \Gamma= -H_B$.
In addition to the chiral symmetry, $H(t)$ preserves the time glide symmetry as
\begin{align}
M_T H(k_x,k_y,t) M_T^\dagger &= H\left(-k_x,k_y,t+\frac T 2 \right), & M_T&=\tau_x.
\end{align}
In particular, $M_T H_A M_T^\dagger= H_B$ holds.

\begin{figure}
\begin{center}
\includegraphics[width=0.8\linewidth]{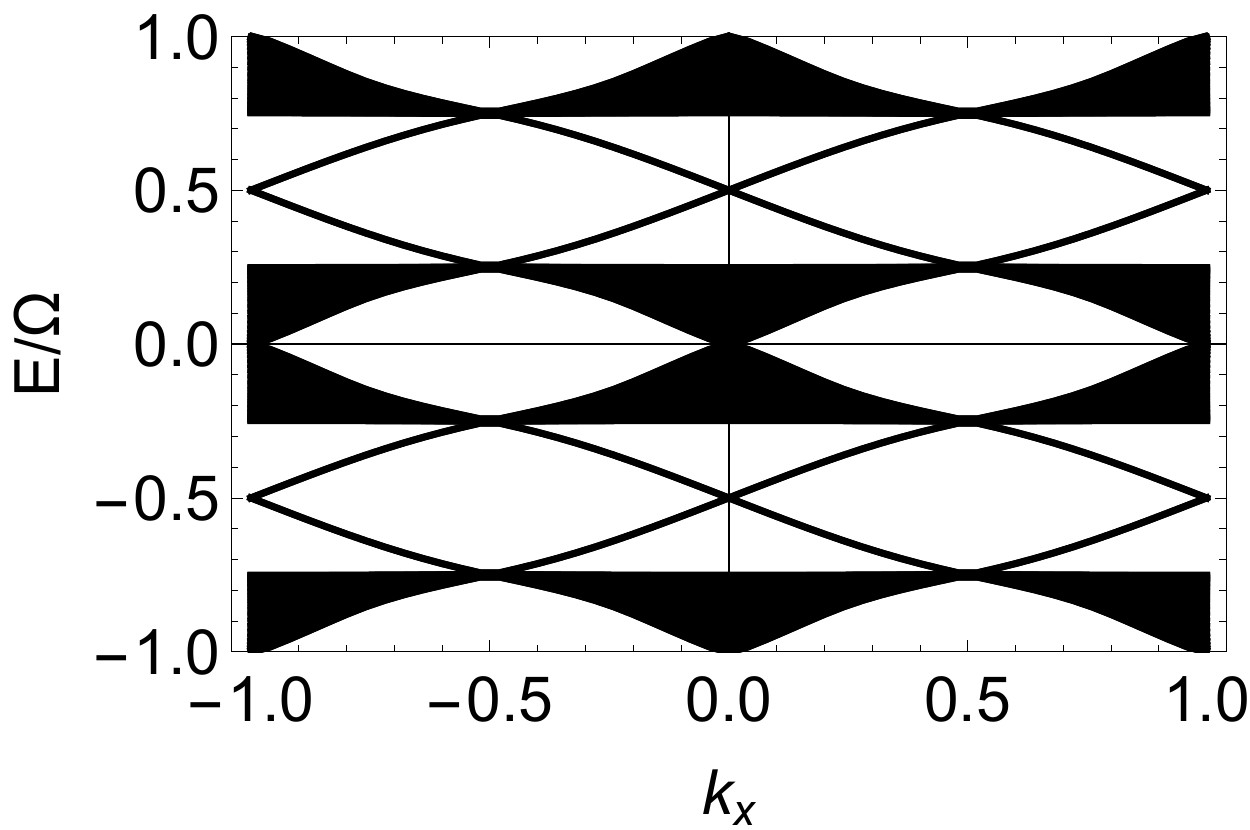}
\end{center}
\caption{Quasienergy spectrum of the 2D model with chiral symmetry and time-glide symmetry. The spectrum is obtained for a cylinder geometry with the open boundary along the $y$ direction with $L_y=20$ sites and the periodic boundary along the $x$ direction.  We used the parameters: $t=1, t'=0.5, \Omega=2$.
Note that  quasienergies in the ``first Brillouin zone'' are $E/\Omega \in [-0.5,0.5]$ .
}
\label{fig: spectrum 2D model}
\end{figure}

Periodically driven systems with time-dependent Hamiltonian H(t) with a period $T$ are described by Floquet Hamiltonians $H_F$ obtained by Fourier transformation along the time-direction as
\begin{align}
(H_F)_{mn}&=
\frac{1}{T} \int_0^T dt e^{i(m-n)\Omega t} H(t) -m \Omega\delta_{mn},
\end{align}
where $\Omega=2\pi/T$ and $m,n$ are Floquet indices.
The eigenvalues of Floquet Hamiltonian are called quasienergy and allow us to capture the dynamics of periodically driven system in terms of a band picture.
The chiral symmetry constrains that the quasienergy spectrum $\epsilon(k_x)$ shows chiral partners at $\epsilon(k_x)$ and $-\epsilon(k_x)$.
In particular, states at $\epsilon=0,\Omega/2$ are special because they can be chiral partners of themselves, which we call chiral zero/$\pi$ modes.
Similarly, the time glide symmetry constrains the spectrum as $\epsilon(k_x)=\epsilon(-k_x)$.

Figure~\ref{fig: spectrum 2D model} shows the quasienergy spectrum of the 2D model $H(t)$,
where we adopted periodic boundary condition along the $x$ direction and the open boundary condition along the $y$ direction.
The spectrum in Fig.~\ref{fig: spectrum 2D model} shows a bulk band around $\epsilon=0$ and a bulk gap around $\epsilon=\Omega/2$, where nontrivial edge states appear. 
This should be contrasted with equilibrium topological phases, which have bulk gap with protected edge states at $\epsilon=0$, and therefore it suggests that the system is in an intrinsically dynamical phase.
Moreover, the edges state appearing in the gap around $\epsilon=\Omega/2$ is protected by the combination of chiral symmetry and time-glide symmetry, since two-dimensional gapped phases with chiral symmetry alone (2D systems in class AIII in the tenfold way) do not support nontrivial phases~\cite{Schnyder08,Roy16}. 
We describe the topological property of these edge states characterized by chiral symmetry and time-glide symmetry in the following.

\subsection{Topological characterization of 2D Floquet phase with chiral symmetry and time glide symmetry}

Before we discuss topological invariant with chiral symmetry and time-glide symmetry, we first review the topological characterization of 1D Floquet topological phases with chiral symmetry (1D systems in class AIII) \cite{Asboth14}.
In order to characterize Floquet topological phases, we study the time-evolution operator given by
\begin{align}
U(  k,t_i \to t_f)&=\mathcal{T}\exp\left[-i\int_{t_i}^{t_f} dt' H( k, t') \right],
\end{align}
since the characterization needs the information of micromotion during the cycle.
By definition, the eigenvalues of the evolution operator $U( k, 0\to T)$ are $e^{-i\epsilon( k) T}$, where $\epsilon( k)$ is the quasienergy.
(We choose $T=1$ for simplicity in the following discussion.)
The chiral symmetry indicates that the time-evolution operator satisfies
\begin{align}
\Gamma U( k, 0\to T/2) \Gamma 
&= U^\dagger( k, T/2 \to T).
\end{align}
When we focus on topological characterization for chiral $\pi$ modes, we can deform the bulk Floquet bands to the quasienergy zero [$\epsilon( k)=0$].
In this case, the evolution operator over the cycle becomes trivial $U( k, 0\to T)=1$ and, in particular,
the equation $U^\dagger( k, T/2 \to T)=U( k, 0\to T/2)$ holds.
If we write the half-period evolution as
\begin{align}
U( k, 0\to T/2)&=
\begin{pmatrix}
a & b \\
c & d \\
\end{pmatrix},
\label{eq: U in chiral basis}
\end{align}
in the basis
\begin{align}
\Gamma&=
\begin{pmatrix}
1 & 0 \\
0 & -1 \\
\end{pmatrix},
\end{align}
this indicates $b=c=0$.
Now we define the winding number $\nu$ for one parameter family of unitary operators $g(k)$ by
\begin{align}
\nu[g(k)]&=\frac{1}{2\pi i}\int dk \, \text{tr} \left( g^\dagger \frac{dg}{dk} \right), 
\end{align}
where $\text{tr}$ denotes a trace over the internal degrees of freedom.
Since $\nu[U( k, 0\to t)]=\nu[U( k, 0\to 0)]=0$ holds due to the continuous deformation $t \to 0$, the two winding numbers of $a(k)$ and $d(k)$ satisfy
$\nu[a]+\nu[d]=0$,
and are not independent.
Thus we can choose the winding number $\nu[d]$ as the topological invariant that characterize chiral $\pi$ modes in 1D class AIII Floquet systems.
In general cases with $U( k, 0\to T)\neq 1$,
the topological invariant for chiral $\pi$ modes $\nu_\pi$ is still given by the winding number as \cite{Asboth14}
\begin{align} 
\nu_\pi&=\nu[d].
\end{align}

Now we proceed to the topological characterization of 2D Floquet phases with chiral symmetry and time glide symmetry.
Since 2D systems in class AIII has no topological number,
we can see that the time-glide symmetry plays a crucial role in characterizing them.
In the presence of time-glide symmetry along the $x$-direction, 1D subsystems at $k_x=0,\pi$ are regarded as 1D class AIII system with an additional $Z_2$ symmetry induced by the time-glide symmetry.
This $Z_2$ symmetry (which we also denote by $M_T$)
is a global $Z_2$ symmetry combined with time translation by the half period, and acts on the Hamiltonian restricted to the 1D subsystem at $k_x=0$ as
\begin{align}
M_T H(0,k_y, t) M_T^{-1}&=H\left(0,k_y, t+\frac T 2\right).
\end{align}
We assume that this subsystem does not have nontrivial winding number $\nu[d]$ in order to exclude the possibility of weak topological insulators of class AIII, where the flat band of chiral $\pi$ mode appears in the quasienergy spectrum. 
Instead, we consider topological invariant similar to the winding number by taking into account the $Z_2$ symmetry $M_T$.
Specifically, under the combination of $\Gamma$ and $M_T$, the Hamiltonian is transformed as
\begin{align}
\Gamma M_T H(0,k_y, t) (\Gamma M_T)^{-1} &=
-H(0,k_y,T/2-t),
\end{align}
which looks similar to the action of $\Gamma$ in Eq.~(\ref{eq: chiral symmetry}) except that the center of time reversal is at $t=T/4$ instead of $t=0$.
This indicates that the time evolution operator satisfies the condition
\begin{align}
\Gamma M_T U \left(0, -\frac T 4 \to \frac{T}{4} \right) (\Gamma M_T)^{-1} 
&=U^\dagger\left(0, -\frac T 4 \to \frac{T}{4}\right)
\label{eq: U(-T/4 ->T/4) and Gamma M_T}
\end{align} 
Therefore, if we write 
\begin{align}
 U\left(0, -\frac T 4 \to \frac{T}{4}\right)&=
\begin{pmatrix}
a' & b' \\
c' & d' \\
\end{pmatrix},
\end{align}
in the basis 
\begin{align}
\Gamma M_T&=
\begin{pmatrix}
1 & 0 \\
0 & -1 \\
\end{pmatrix},
\end{align}
in a similar way to Eq.~(\ref{eq: U in chiral basis}),
we can define the winding number $\nu[d']$ for the 1D subsystem at $k_x=0$. 
Similarly, we can define the winding number $\nu[d']$ at $k_x=\pi$.
Nonzero winding numbers $\nu[d']$ result in the presence of chiral $\pi$ modes at glide symmetric points ($k_x=0,\pi$).
While the presence of chiral $\pi$ mode is not protected away from the glide symmetric points, the continuity of quasienergy spectrum with $k_x$ ensures the presence of an edge state within the bulk gap.
Thus nonzero winding numbers $\nu[d']$ at time glide planes characterize the nontrivial edge states in the $\pi$ gap.

The above topological number $\nu[d']$ characterizing 2D Floquet topological phases with $\Gamma$ and $M_T$ requires anticommutation relation between $\Gamma$ and $M_T$.
To see this, we show that if $[\Gamma,M_T]=0$, the winding number vanishes identically ($\nu[d']=0$):
If $\Gamma M_T$ and $M_T$ commute, they  can be simultaneously diagonalized as
\begin{align}
\Gamma M_T&=
\begin{pmatrix}
1 & 0 \\
0 & -1 \\
\end{pmatrix},
 & 
M_T&=
\begin{pmatrix}
M_T^+ & 0 \\
0 & M_T^- \\
\end{pmatrix}.
\end{align}
By combining this equation and Eq.~(\ref{eq: U(-T/4 ->T/4) and Gamma M_T}),
we obtain $M_T^- d' (M_T^-)^{-1}=d'^\dagger$. 
Since the winding number satisfies the relationships $\nu[g]=\nu[g'gg'^{-1}]$ and $\nu[g]=-\nu[g^\dagger]$, these lead to $\nu[d']=-\nu[d']=0$ when $[\Gamma,M_T]=0$.
Thus the nonzero winding number requires the anticommutation relation $\{\Gamma, M_T\}=0$.
These structures are concisely captured through a Clifford algebra analysis, which will be detailed in Sec.~\ref{sec: classification time glide}. 

The anticommutation relationship between the chiral symmetry and the time glide symmetry shows that the time glide symmetric Floquet topological phase is distinct from conventional topological crystalline insulators (TCIs). It is known that nontrivial TCIs exist in 2D class AIII with (static) reflection symmetry \cite{Chiu13,Morimoto13,Shiozaki14}. A natural question is whether the time glide symmetric Floquet topological insulators can be obtained by deforming a conventional TCI with a time glide symmetric perturbation. The condition $\{\Gamma, M_T\}=0$ shows that this is not the case because nontrivial TCIs requires the commutation relationship $[\Gamma, R_x]=0$ ($R_x$ being a reflection symmetry along the $x$ direction). Thus perturbing a TCI by relaxing static $R_x$ into dynamical $M_T$ cannot result in a time glide symmetric Floquet topological phase, which clarifies these two phases are distinct sets of topological phases.

Finally, we explicitly calculate the winding number $\nu[d']$ for the 2D model given by Eq.~(\ref{eq: 2d model}).
We focus on the mirror invariant subspaces at $k_x=0,$ or $k_x=\pi$ (which give the same topological numbers as we will see below).
Since we can deform the interlayer hopping $t'$ to zero without closing the bulk gap, we can focus on two chains coupled with complex hoppings described by 
\begin{align}
H(k_y,t)=\sigma_x \cos \frac{k_y}{2} + \eta(t)\sigma_y \tau_y \sin \frac{k_y}{2},
\end{align}
where we also set $2t=1$ for simplicity.
In this case, the time evolution operator $U(-T/4\to T/4)$ can be obtained by setting $T=2\pi$ as
\begin{align}
U(-T/4 \to T/4)&=\exp \left[-i \frac T 4 \left(\sigma_x \cos \frac{k_y}{2} - \sigma_y \tau_y \sin \frac{k_y}{2} \right)\right] \n
&\quad \times \exp\left[-i \frac T 4\left(\sigma_x \cos \frac{k_y}{2} + \sigma_y \tau_y \sin \frac{k_y}{2} \right)\right] \n
&=-\cos(k_y)-i \sin(k_y) \sigma_z \tau_y.
\end{align}
Since $\Gamma M_T=\sigma_z\tau_y$,
the topological number for the mirror invariant subspace is given by 
\begin{align}
\nu[d']&=\nu[-e^{-i k_y} \tensor \openone_2 ]=-2,
\end{align}
where $\openone_2$ denotes a 2 by 2 identity matrix in the $\sigma_z \tau_y=-1$ sector.
This nonzero topological number defined with time glide symmetry protects the two fold degeneracy at $E=\Omega/2$ in the mirror invariant subspace, and hence, the gap closings at $k_x=0,\pi$ shown in Fig.~\ref{fig: spectrum 2D model}.

\subsection{3D model of class A and time-glide \label{subsec: 3D A}}

\begin{figure}
\begin{center}
\includegraphics[width=\linewidth]{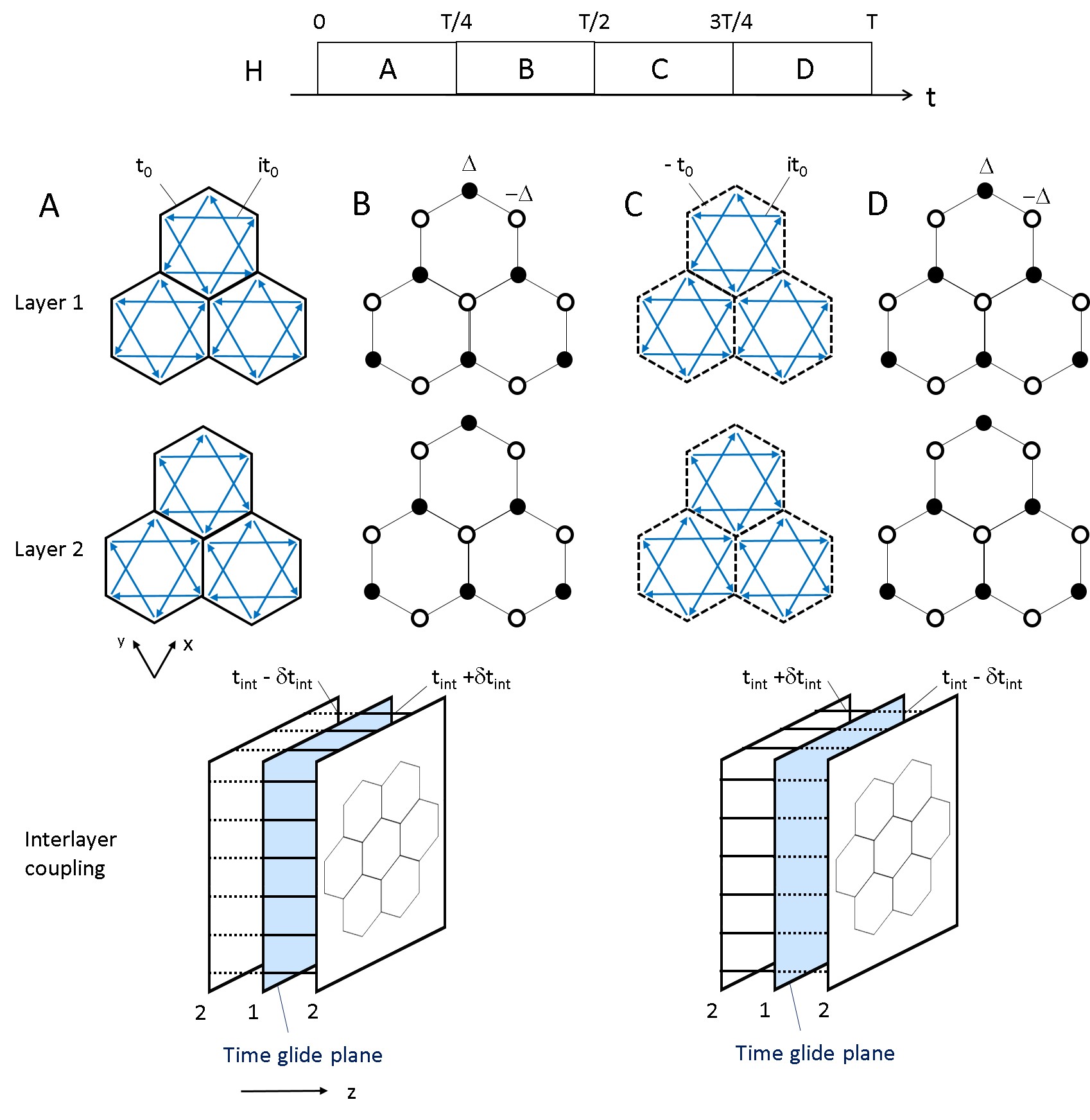}
\end{center}
\caption{Schematic picture of the 3D model with time-glide symmetry.
The model consists of a four step drive, denoted by A, B, C and D.
Solid and dotted lines represent static hoppings with $t$ and $t'$, respectively.
Blue arrows represent alternating hoppings with the amplitude $it$ ($-it$) along (opposite to) the direction of the arrow.
}
\label{fig: 3D A+Tglide}
\end{figure}

\begin{figure}
\begin{center}
\includegraphics[width=\linewidth]{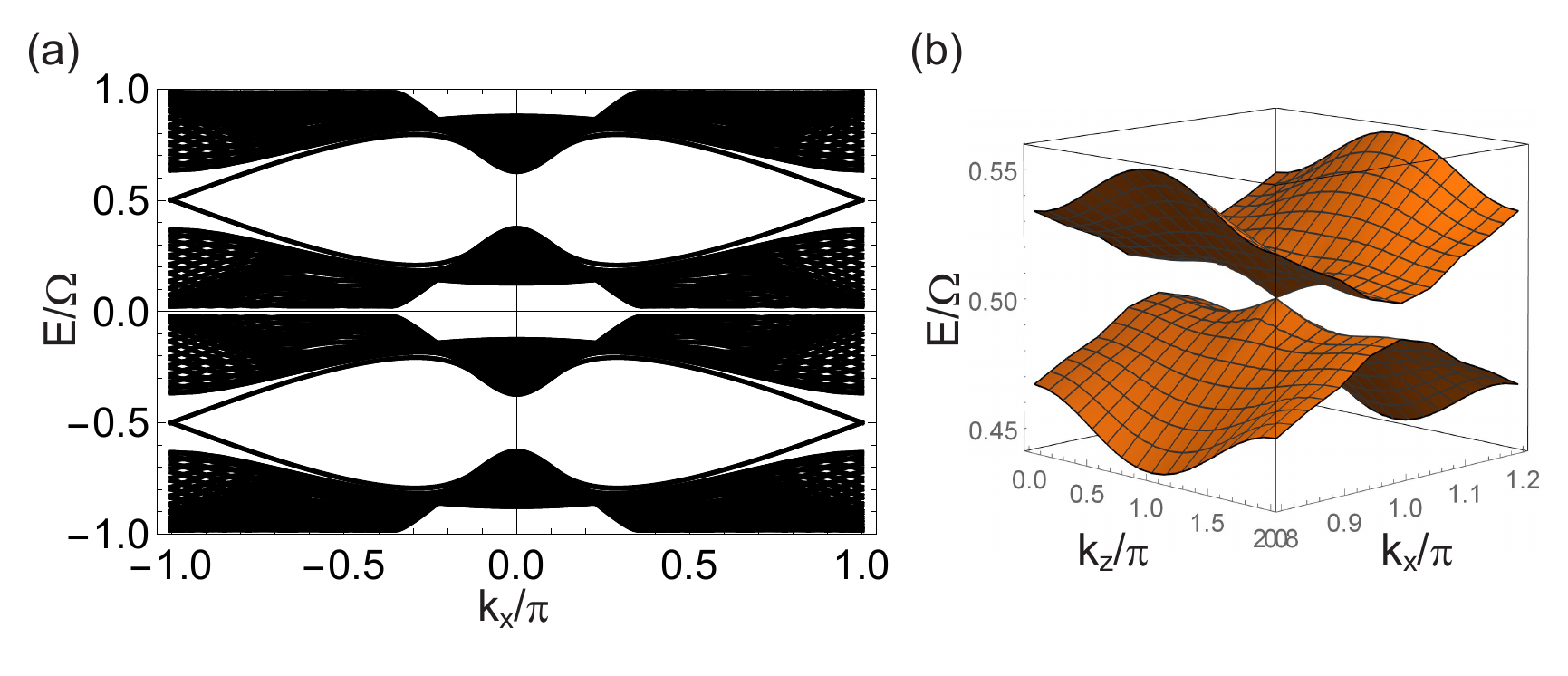}
\end{center}
\caption{Quasienergy spectrum of the 3D model with time-glide symmetry. We consider a slab model which has periodic boundaries along the $x$ and $z$ directions and open boundary along the $y$ direction.
(a) Quasienergy spectrum at $k_z=0$ which is time glide symmetric plane. 
(b) Energy dispersion of the surface state that appears within the $\pi$ gap.
We used the parameters: $t_0=0.25, t_\t{int}=0.04, \delta t_\t{int}=0.02, \Delta=1, \Omega=1$ and $L_y=30$..
}
\label{fig: spectrum 3D A}
\end{figure}

Next, we present a 3D model that supports a Floquet topological phase protected by the time-glide symmetry. We focus on an insulating phase with charge U(1) symmetry which belongs to the symmetry class A in the AZ classification.
The 3D model we study is defined on a stack of the honeycomb lattice and is described by the time-dependent Hamiltonian 
$
H(t)= H_\t{intra}(t) + H_\t{inter}(t)
$ which is given as follows.
First, the intralayer part of the Hamiltonian $H_\t{intra}$ is given by the four step driving by the Haldane model \cite{Haldane} and staggered potential for honeycomb lattice as
\begin{align}
H_\t{intra}(t)&=
\begin{cases}
H_A , & (0 \le t < \frac{T}{4}) \\
H_B , & (\frac{T}{4} \le t < \frac{T}{2}) \\
H_C , & (\frac{T}{2} \le t < \frac{3T}{4}) \\
H_D . & (\frac{3T}{4} \le t < T) \\
\end{cases}
\label{eq: H intra 3D}
\end{align}
with
\begin{align}
H_A&=
t_0 \sum_{m,\langle i,j \rangle} c_{m,i}^\dagger c_{m,j} 
+t_0 \sum_{m,\langle \langle i,j \rangle \rangle} (-1)^m e^{i\Phi_{ij}} c_{m,i}^\dagger c_{m,j} , \\
H_C&=
-t_0 \sum_{m,\langle i,j \rangle} c_{m,i}^\dagger c_{m,j}  
+t_0 \sum_{m,\langle \langle i,j \rangle \rangle} (-1)^m e^{i\Phi_{ij}} c_{m,i}^\dagger c_{m,j} , \\
H_B&=H_D=\Delta \sum_{m,i} \eta_i c_{m,i}^\dagger c_{m,i} ,
\end{align}
where 
$c_{m,i}$ is the annihilation operator of an electron with the layer index $m$ and the site index $i$-th within the honeycomb lattice,
$\langle i,j \rangle$ denotes the nearest neighbor sites $i$ and $j$,
$\langle \langle i,j \rangle \rangle$ denotes the next nearest neighbor sites, the phase factor $e^{i\Phi_{ij}}=\pm i$ according to the direction of arrows in Fig.~\ref{fig: 3D A+Tglide},
and $\eta_i=\pm 1$ for two distinct sublattices of the honeycomb lattice. 
Next, the interlayer part of the Hamiltonian is given by
\begin{align}
&H_\t{inter}= \n
&\begin{cases}
\sum_{m,i} [t_\t{int} + (-1)^i \delta t_\t{int}]c_{m+1,i}^\dagger c_{m,i} + \t{h.c.} & (0\le t< \frac T 2) \\
\sum_{m,i} [t_\t{int} - (-1)^i \delta t_\t{int}]c_{m+1,i}^\dagger c_{m,i} + \t{h.c.} & (\frac T 2 \le t< T) \\
\end{cases}
\end{align}
In the momentum space (where $k_x$ and $k_y$ are measured along the directions specified in Fig.~\ref{fig: 3D A+Tglide}),
the Hamiltonian is written in the 4 by 4 form as
\begin{align}
H_\t{intra}&=
\begin{cases}
t_0 \Vec d_\t{H} \cdot \Vec \sigma, & (0 \le t < \frac{T}{4}) \\
\Delta \sigma_z, & (\frac{T}{4} \le t < \frac{T}{2}) \\
t_0 \sigma_z (\Vec d_\t{H} \cdot \Vec \sigma) \sigma_z, & (\frac{T}{2} \le t < \frac{3T}{4}) \\
\Delta \sigma_z, & (\frac{3T}{4} \le t < T) \\
\end{cases} 
\end{align}
with 
$
\Vec d_\t{H}
= \big( 1+ \cos k_x + \cos k_y, \sin k_x + \sin k_y, [-2 \sin k_x +2 \sin k_y + 2 \sin(k_x-k_y)]\tau_z \big),
$
and
\begin{align}
H_\t{inter} &= 
\begin{cases}
t_\t{int} \cos \frac{k_z}{2} \tau_x + \delta t_\t{int} \sin \frac{k_z}{2} \tau_y, & (0 \le t < \frac T 2) \\
t_\t{int} \cos \frac{k_z}{2} \tau_x - \delta t_\t{int} \sin \frac{k_z}{2} \tau_y, & (\frac T 2 \le t < T) \\
\end{cases}
\end{align}
where $\sigma_i$ and $\tau_i$ are Pauli matrices acting on sublattice and layer degrees of freedom.
This model has a time-glide symmetry with $x$-$y$ plane as the mirror plane. More explicitly, one finds $M_T H(t) M_T^{-1}=H(t+T/2)$, with the time-glide operator given in this 4 by 4 representation by
\begin{align}
M_T &= \sigma_z 
e^{-i(1-\tau_z) k_z/2}.
\end{align}

The quasienergy spectrum for the above 3D model is shown in Fig.~\ref{fig: spectrum 3D A}.
We consider the system with the open boundary condition along the $y$-direction and the periodic boundary condition along the $x$ and $z$ directions. The quasienergy spectrum shows the bulk gap around $\pm \Omega/2$ and there appears a surface state.
The inset of Fig.~\ref{fig: spectrum 3D A} shows a blowup of the surface state and shows that a Dirac fermion with linear dispersion is realized at the surface around $(k_x,k_z)=(\pi,\pi)$.
We note that the gapless point is located at one of the mirror invariant plane with $M_T$ at $k_z=0,\pi$.
While 3D systems in the symmetry class A do not possess any topological number according to the ten-fold way classification \cite{Schnyder08,Ryu10,Kitaev09},
this Dirac surface state is protected by the time-glide symmetry, as we now explain.


The topological number for the 3D systems with time-glide symmetry is defined by focusing on the mirror invariant plane. In the case of time glide along the $z$ direction in the above,
we focus on the mirror invariant plane at $k_z^0=0,\pi$ and the restricted 2D Hamiltonian having global $Z_2$ symmetry combined with half time translation
\begin{align}
M^T H(k_x,k_y,k_z^0,t)(M^T)^{-1} &= H(k_x,k_y,k_z^0,t+T/2).
\end{align}
Topological numbers are defined for this effective 2D system as follows.

We define a topological number of a 2D system with global $Z_2$ symmetry $g_T$ combined with half time translation.
First we notice that the time evolution for the full period is described by that for the half period as
\begin{align}
U(k_x,k_y,0\to T)&=g_T U_h g_T U_h, \\
U_h(k_x,k_y)&=U \left(k_x,k_y,0 \to \frac T 2 \right).
\label{eq: def of Uf}
\end{align}
Here we used a representation for a order-two unitary symmetry $g_T$ such that $g_T=g_T^{-1}$.
Since the Floquet Hamiltonian is obtained from the time evolution operator,
the quasienergy spectrum is determined by eigenvalues of $U_h$ as
\begin{align}
H_F(k_x,k_y)&=\frac{i}{T} \ln U(k_x,k_y,0\to T) = \frac{i}{T} \ln[(g_T U_h)^2].
\end{align}
(For simplicity, we choose $T=1$ hereafter.)
Now we assume that the bulk quasienergy spectrum has a gap at $E=\pi$.
In this case, we can continuously deform the quasienergy to $E=0$ for the entire BZ. This means the eigenvalues of $g_T U_h$ are adiabatically connected to $\pm 1$ without closing the gap at $E=\pi$. 
For trivial time evolution $U_h=1$, eigenvalues of $g_T U_h$ are given by those of $g_T$.
Let us focus on two band system and the time evolution operators belong to $U(2)$ (where we may drop the $U(1)$ part for simplicity, without loss of generality, and focus on the $SU(2)$ part). Suppose that $g_T$ has two eigenvalues $+1$ and $-1$ (e.g., $g_T=\sigma_z$).
If the eigenvalues of $g_T U_h$ are both $+1$ or both $-1$, the system is topologically distinct from trivial time evolution. This is because deforming $U_h$ to 1 requires that one of the eigenvalues of $g_T U_h$ continuously changes from $-1$ to $+1$ (or from $+1$ to $-1$) as a $U(1)$ variable, and in this process the eigenvalue of $g_T U_h$ passes $\pm i$ where the bulk $\pi$ gap of $U(k_x,k_y,0\to T)$ is closed.
Thus the system is nontrivial if passing the points $U_h= \pm i g_T$ cannot be avoided in the deformation into the trivial evolution $U_h=1$.
This situation is achieved when $U_h$ as an $SU(2)$ operator wraps around the points either $ig_T$ or $-ig_T$. This prevents trivialization of $U_h$ into 1.
In the case of general number of bands,
this obstruction is described by the Chern number for $g_T U_h$ as follows.
First we consider time evolution operators after spectral flattening $U(k_x,k_y,0\to T)=1$. This is possible because we can find $g_T$ symmetric deformation of $U(t)$ to $U(T)=(g_T U_h)^2=1$.
In this case, Eq.~(\ref{eq: def of Uf}) indicates that the half time evolution satisfies
\begin{align}
g_T U_h &=(g_T U_h)^\dagger,
\end{align}
and the operator $g_T U_h$ becomes hermitian.
Since the hermitian operator $g_T U_h(k_x,k_y)$ has two parameters in the 2D system and eigenvalues of $g_T U_h(k_x,k_y)$ are $\pm 1$ (having a spectral gap), we can define a Chern number for $g_T U_h$.
This Chern number coincides with the wrapping number defined in the above for the two band system since the Chern number of $g_T U_h$ means a nontrivial wrapping around the identity operator, and hence, nontrivial wrapping of $U_h(k_x,k_y)$ around $g_T$. (We note that $ig_T$ is replaced with $g_T$ here, because the energy gap is located at $E=0$ rather than $E=\pi$.)


\begin{figure}
\begin{center}
\includegraphics[width=\linewidth]{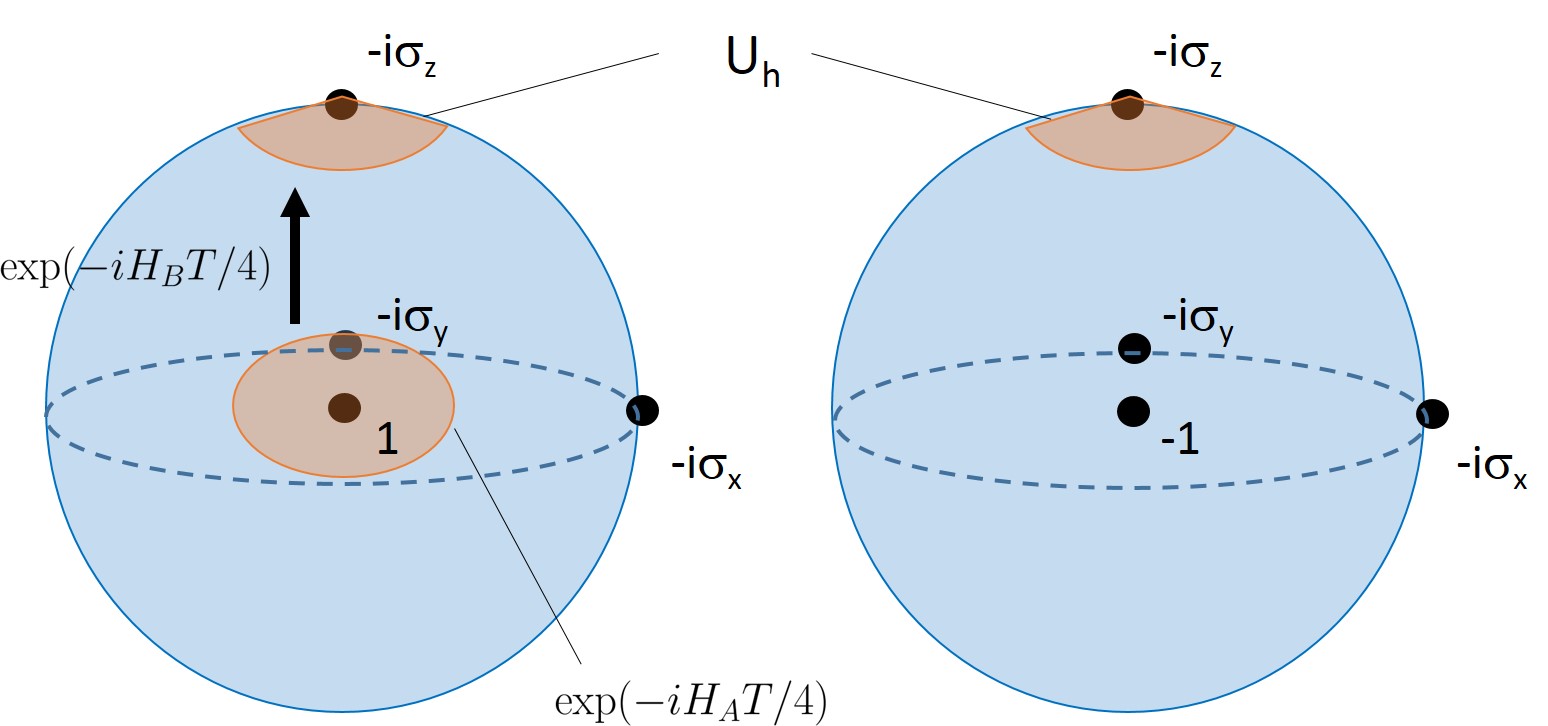}
\end{center}
\caption{Schematic picture of the topological number at the mirror invariant plane. Time evolution operators for $H_\t{intra}$ for each layer belongs to SU(2) which is described by $S^3$. The sphere $S^3$ is illustrated by two solid balls whose surfaces are identified. Nontrivial topological phase is characterized by the wrapping of $\pm i g_T$ by the half time evolution operator $U_h$. 
Since closing of the gap at $E=\pi$ takes place at $U_h=\pm i g_T$, the topological number is defined as a wrapping number around these points. After spectral flattening, $U_h$ is deformed within a disk ($S^2$) containing $1, -1$ and the blue dotted circle, and this wrapping number coincides with the Chern number defined for $g_T U_h$.
}
\label{fig: 3D topological invariant}
\end{figure}

Finally we show that the 3D model with time glide symmetry has a nontrivial topological number defined in the above.
We focus on the mirror invariant plane at $k_z^0=0,\pi$. Since we can deform the interlayer coupling $H_\t{inter}$ to zero without closing the bulk gap, we consider the case of no interlayer coupling $t_\t{int}=\delta t_\t{int}=0$.
In this case the Chern number of $g_T U_h$ in the mirror invariant plane can be obtained by computing those for two kind of decoupled layers $\tau_z=\pm 1$.
Let us focus on the layer 1 with $\tau_z=+1$, where we have a two band system and we can deduce the topological number from wrapping number of $ig_T$ by $U_h$ within $U(2)$. Since the intralayer Hamiltonian $H_\t{intra}$ is traceless (i.e., consisting of sums of three Pauli matrices), the half time evolution $U_h$ belongs to the $SU(2)$ part (which can be visualized by $S^3$ as in Fig.~\ref{fig: 3D topological invariant}), and we can reduce the topological characterization of $U_h$ to whether $U_h$ winds around the point $i\sigma_z=ig_T$ in $SU(2)$.
The half time evolution for Eq.~(\ref{eq: H intra 3D}) is written as
\begin{align}
U_h=\exp(-iH_B T/4)\exp(-iH_A T/4).
\end{align}
The first step of the driving $\exp( -i H_A T/4) $ wraps around the identity element 1. This is because $H_A=\Vec d_H \cdot \Vec \sigma$ is the Hamiltonian of the Haldane model which has a nonzero Chern number; the vector $\Vec d_H$ wraps around the origin, and hence, its exponential map $\exp( -i H_A T/4)$ winds around the identity element 1 when the magnitude $t_0 T/4$ is small which  is the case for the parameters in Fig. \ref{fig: spectrum 3D A}. 
By setting $\Delta T/4=\pi/2$, the next step leads to the factor 
$\exp(-iH_B T/4)=-i\sigma_z$ which maps the identity element 1 to the point $-i \sigma_z$. Thus $\exp( -i H_A T/4)$ wrapping around 1 is mapped to $U_h$ wrapping around $-i\sigma_z$, which ensures a topologically nontrivial configuration with global $Z_2$ symmetry $g_T$ with half time translation.  Once we perform the spectral flattening for the Floquet operator $U(T)$, the Chern number of $\sigma_z U_h$ is nonzero, say 1 in this case.
In a similar way, the layer 2 ($\tau_z=-1$) gives the Chern number $-1$ for $\sigma_z U_h$.
At the mirror invariant plane $k_z=0$, these two contributions add up because of $M_T=\sigma_z$ and give zero Chern number. 
In contrast, at the mirror invariant plane $k_z=\pi$, the difference of these two contributions is the topological number because of $M_T=\sigma_z \tau_z$, which gives the nonzero Chern number 2.
This is consistent with the band structure in Fig.~\ref{fig: spectrum 3D A} where the gapless surface state appears at $k_z=\pi$ where the bulk topological number with the glide symmetry becomes nonzero.

\begin{table*}[tb]
\begin{center}
\caption{\label{table: AZ classes}
The ten ten-fold way classification for noninteracting Floquet topological phases.
Two complex and eight real symmetry classes are characterized
by the presence or the absence of 
time-reversal symmetry ($T$), 
particle-hole symmetry ($C$), 
and chiral symmetry ($\Gamma$).
Their presence is complemented
by the sign multiplying the identity 
in $T^{2}=\pm1$ or $C^{2}=\pm1$, 
and by $1$ for $\Gamma$.
Their absence is indicated by 0.
For each spatial dimension $d$, nontrivial topological phases are characterized by $\mathbb{Z}$, $\mathbb{Z}_2$ topological numbers. 
For symmetry classes without PHS or chiral symmetry, $n$ denotes the number of gaps in the quasienergy spectrum.
\\
}
\begin{tabular}[t]{ c c c c c c c c c c c c }
\hline \hline
Class 
& 
$T$ 
& 
$C$ 
& 
$\Gamma$ 
& 
$d=0$
& 
$d=1$ 
&
$d=2$ 
&
$d=3$ 
&
$d=4$
& 
$d=5$ 
&
$d=6$ 
&
$d=7$ 
\\
\hline
A   
& 
0 
& 
0    
&  
0     
& 
$\mathbb{Z}^n$   
& 
0
&
$\mathbb{Z}^n$   
&
0
& 
$\mathbb{Z}^n$   
& 
0
&
$\mathbb{Z}^n$   
&
0
\\
AIII  
& 
0 
& 
0    
&  
1     
& 
0 
&
$\mathbb{Z}^2$   
&
0
&
$\mathbb{Z}^2$ 
& 
0 
&
$\mathbb{Z}^2$   
&
0
&
$\mathbb{Z}^2$ 
\\
\hline
AI   
& 
$+1$ 
& 
0    
&  
0     
& 
$\mathbb{Z}^n$
&
0
&
0
&
0
&
$\mathbb{Z}^n$
&
0
&
$\mathbb{Z}_{2}^n$
&
$\mathbb{Z}_{2}^n$
\\
BDI  
& 
$+1$ 
& 
$+1$ 
&  
1     
& 
$\mathbb{Z}_{2}^2$ 
&
$\mathbb{Z}^2$
&
0
&
0
&
0
&
$\mathbb{Z}^2$
&
0
&
$\mathbb{Z}_{2}^2$
\\
D    
& 
0    
& 
$+1$ 
&  
0     
& 
$\mathbb{Z}_{2}^2$ 
&
$\mathbb{Z}_{2}^2$ 
&
$\mathbb{Z}^2$
&
0
&
0
&
0
&
$\mathbb{Z}^2$
&
0
\\
DIII 
& 
$-1$ 
& 
$+1$ 
&  
1     
& 
0              
&
$\mathbb{Z}_{2}^2$ 
&
$\mathbb{Z}_{2}^2$ 
&
$\mathbb{Z}^2$
&
0
&
0
&
0
&
$\mathbb{Z}^2$
\\
AII  
& 
$-1$ 
& 
0    
&  
0     
& 
$\mathbb{Z}^n$   
&
0
&
$\mathbb{Z}_{2}^n$ 
&
$\mathbb{Z}_{2}^n$ 
&
$\mathbb{Z}^n$
&
0
&
0
&
0
\\
CII  
& 
$-1$ 
& 
$-1$ 
&  
1     
& 
0              
&
$\mathbb{Z}^2$
&
0
&
$\mathbb{Z}_{2}^2$
&
$\mathbb{Z}_{2}^2$ 
&
$\mathbb{Z}^2$
&
0
&
0
\\
C    
& 
0    
& 
$-1$ 
&  
0     
& 
0              
&
0
&
$\mathbb{Z}^2$
&
0
&
$\mathbb{Z}_{2}^2$
&
$\mathbb{Z}_{2}^2$
&
$\mathbb{Z}^2$
&
0 
\\
CI   
& 
$+1$
& 
$-1$ 
&  
1     
& 
0              
&
0
&
0
&
$\mathbb{Z}^2$
&
0
&
$\mathbb{Z}_{2}^2$ 
&
$\mathbb{Z}_{2}^2$ 
&
$\mathbb{Z}^2$ 
\\
\hline \hline
\end{tabular}
\end{center}
\end{table*}

\section{Classification of Floquet topological phases}
Having provided concrete examples for time-protected Floquet topological phases in 2 and 3D, we now classify noninteracting Floquet topological phases with time glide symmetry in all symmetry classes in arbitrary dimensions. To this end, we use classification theory of topological insulators based on Clifford algebras~\cite{Morimoto13} and apply it to Floquet topological phases. We also consider topological phases realized with a ``time-screw'' symmetry, which corresponds to a  two-fold rotation together with a  half-period translation. 

\subsection{Tenfold way classification of Floquet topological phases}
Before proceeding to the classification theory of Floquet topological phases with time glide symmetry,
we review the ten-fold way classification theory for Floquet topological phases~\cite{Roy16}, which serves as a basis for studying the cases with time glide symmetry.
Topological characterization of Floquet topological phases involves data of time evolution operators for whole period, i.e., $U(t)$ with $t\in [0,T)$, rather than just Floquet operator $U(t=T)$ (or equivalently, the Floquet Hamiltonian $H_F$).
In order to study topological properties of the family of unitary operators $U(t)$, we instead study an effective Hamiltonian that is made of $U(t)$ which enables us to apply classification technique developed for equilibrium topological phases~\cite{Schnyder08,Ryu10,Kitaev09,Morimoto13}, as we will explain below.

We consider the symmetrized time-evolution operator
\begin{align}
U_S(\bm k,t)&=\mathcal{T}\exp \left[-i\int_{\frac{T-t}{2}}^{\frac{T+t}{2}} dt' H(\bm k, t') \right]
\n
&\equiv \lim_{N\to\infty} \prod_{n=0}^N \left(1-i\frac{t}{N}H\left(\bm k, \frac{T-t}{2}+ \frac{n t}{N} \right)\right),
\end{align}
where $\mathcal{T}$ denotes the time-ordering.
This family of unitary operator $U_S(\bm k, t)$ encodes topological data of Floquet topological phases (and has topological data equivalent  to usual time evolution operators $U(\bm k,t)$).
This unitary operator satisfies $U_S(\bm k, -t)=U_S^\dagger(\bm k, t)$.
In addition, we assume that the operator $U_S(\bm k,t)$
 satisfies $U_S(\bm k, 0)=U_S(\bm k, T)=1$ for the Floquet period $T$ by an appropriate deformation of the time-dependent Hamiltonian. This condition is equivalent to considering spectral-flattened Floquet Hamiltonian $H_F=0$,
and it is satisfied if we require the existence of a gap in the quasienergy spectrum around $\Omega/2$.

Instead of studying the unitary operator $U_S(\bm k, t)$ itself, we consider a Hamiltonian given by
\begin{align}
H_S(\bm k, t)=
\begin{pmatrix}
0 & U_S(\bm k, t) \\
U_S^\dagger(\bm k, t) & 0
\end{pmatrix},
\label{eq: H_S}
\end{align}
which satisfies $H_S^2=1$.
This Hamiltonian is smoothly defined for $(\bm k, t)\in T^d \times S^1$ and encodes the topological nature of the periodically-driven system.

The symmetry constraints for the original time-dependent Hamiltonian $H(\bm k,t)$ result in those for $H_S(\bm k, t)$.
We consider time-reversal, particle-hole, and chiral symmetries 
(denoted by $T,C,\Gamma$, respectively) according to the ten-fold way classification of topological insulators:
\begin{align}
TH(\bm k, t)T^{-1}=H(-\bm k, -t), \\
CH(\bm k, t)C^{-1}=-H(-\bm k, t), \\
\Gamma H(\bm k, t)\Gamma^{-1}=-H(\bm k, -t),
\end{align} 
where $T,C$ are antiunitary ($\{T,i\}=\{C,i\}=0$) 
and $\Gamma$ is unitary ($[\Gamma,i]=0$).
We assume that $T,C$ and $\Gamma$ commute with each other, if they are present, without loss of generality.
Note that the chiral symmetry changes the sign of $t$ (to $-t$) because it is given by $TC$ when both $T$ and $C$ exist.
Accordingly, the time-evolution operator satisfies
\begin{align}
TU_S(\bm k, t)T^{-1}
&=
T \prod_{n=0}^N \left[1-i \frac{t}{N} H \left(\bm k, \frac{T-t}{2}+\frac{nt}{N} \right) \right]T^{-1}
\n
&=\prod_{n=0}^N \left[1+i \frac{t}{N} H \left(-\bm k, \frac{-T+t}{2}-\frac{nt}{N} \right) \right]
\n
&=\prod_{n=0}^N \left[1-i \frac{-t}{N} H \left(-\bm k, \frac{T-(-t)}{2}+\frac{-nt}{N} \right) \right]
\n
&=U_S(-\bm k, -t)=U^\dagger(-\bm k, t),
\end{align}
\begin{align}
CU_S(\bm k, t)C^{-1}
&=
C \prod_{n=0}^N \left[1-i \frac{t}{N} H\left(\bm k, \frac{T-t}{2}+\frac{nt}{N} \right) \right]C^{-1}
\n
&=\prod_{n=0}^N \left[1-i \frac{t}{N} H\left(-\bm k, \frac{T-t}{2}+\frac{nt}{N} \right)\right]
\n
&=U_S(-\bm k, t),
\end{align}
\begin{align}
\Gamma U_S(\bm k, t)\Gamma^{-1}
&=
\Gamma \prod_{n=0}^N \left[1-i \frac{t}{N} H \left(\bm k, \frac{T-t}{2}+\frac{nt}{N} \right)\right]\Gamma^{-1}
\n
&=\prod_{n=0}^N \left[1+i \frac{t}{N} H \left(\bm k, \frac{-T+t}{2}-\frac{nt}{N} \right)\right]
\n
&=\prod_{n=0}^N \left[1+i \frac{t}{N} H\left(\bm k, \frac{T-t}{2}+\frac{(N-n)t}{N} \right)\right]
\n
&=U_S^\dagger(\bm k, t),
\end{align}
These relations lead to symmetry constraints for $H_S$ given by
\begin{align}
T'H_S(\bm k, t)T'^{-1}&=H_S(-\bm k, t), & T'=T\tensor \sigma_x, \\
C'H_S(\bm k, t)C'^{-1}&=H_S(-\bm k, t), & C'=C\tensor \sigma_0, \\
\Gamma' H_S(\bm k, t)\Gamma'^{-1}&=H_S(\bm k, t), & \Gamma'=\Gamma\tensor \sigma_x,
\end{align}
where $\sigma_i$ are Pauli matrices acting on two by two matrix degrees of freedom of $H_S$.
In addition, $H_S$ satisfies an inherent chiral symmetry as
\begin{align}
\tilde \Gamma H_S(\bm k, t) \tilde \Gamma^{-1}&=-H_S(\bm k, t), & \tilde \Gamma=1 \tensor \sigma_z.
\end{align}

Classification of  Floquet topological phases is obtained by studying  topological characters of $H_S(\bm k,t)$. The effective Hamiltonian $H_S(\bm k,t)$ can be classified in a similar manner to equilibrium topological phases. Specifically, we can map the classification problem of $H_S(\bm k,t)$ to that for $d+1$D TIs or line defects in $d$D systems as detailed in Appendix~\ref{app: tenfold way}. The obtained classification table is shown in Table~\ref{table: AZ classes}.
It shows that 
Floquet topological phases share the same topological classification with equilibrium topological phases.
One remarkable feature is that topological numbers can be defined for each energy gap in the quasienergy spectrum. 
When particle-hole symmetry (PHS) or chiral symmetry is present, we focus on gaps including quasienergy 0 and $\pi$ which are particle-hole symmetric, hence, we can define two topological numbers.

\begin{table*}[tb]
\begin{center}
\caption{\label{table: extension tenfold way}
Clifford algebras characterizing the Floquet topological phases.
Clifford algebras consist of mass and kinetic gamma matrices and symmetry operators. The space of possible Dirac masses $V$ is obtained from the extension problem of the Clifford algebra with respect to the mass term $\gamma_0$. The zeroth homotopy group of $V$ gives the Abelian group that characterizes Floquet topological phases. In the last column, we show Abelian groups $\pi_0(V)$ for the cases of zero dimensional systems. The periodic structure with the dimensions follows from the shift in the space of Dirac masses $V$ with $d$. 
\\
}
\begin{tabular}[t]{ c c c c c }
\hline \hline
~Class~ 
& 
~Clifford algebras~
& 
~Extension problem~
& 
~Space of masses $V$~
&
~$\pi_0(V)_{d=0}$~
\\
\hline
A   
& 
$\{ \gamma_0, \gamma_1, \ldots, \gamma_d, \gamma_t, \tilde \Gamma \}$
&
$Cl_{d+2} \to Cl_{d+3}$
&
$C_d$ & $\mathbb{Z}$
\\
AIII
&
$\{ \gamma_0, \gamma_1, \ldots, \gamma_d, \gamma_t, \tilde \Gamma, \tilde \Gamma \Gamma' \}$
&
$Cl_{d+3} \to Cl_{d+4}$
&
$C_{d+1}$ & $0$
\\
\hline
AI   
&
$\{ i\gamma_0, i\gamma_t; T', iT', \gamma_1, \ldots, \gamma_d, \tilde \Gamma \}$
&
$Cl_{1,d+3} \to Cl_{2,d+3}$
&
$R_{-d}$ & $\mathbb{Z}$
\\
BDI 
& 
$\{ i\gamma_0, i\gamma_t, \tilde \Gamma \Gamma' ; T', iT', \gamma_1, \ldots, \gamma_d, \tilde \Gamma \}$
&
$Cl_{2,d+3} \to Cl_{3,d+3}$
&
$R_{1-d}$ & $\mathbb{Z}_2$
\\
D & 
$\{ i\gamma_0, i\gamma_t, i\tilde \Gamma; C', iC', \gamma_1, \ldots, \gamma_d \}$
&
$Cl_{2,d+2} \to Cl_{3,d+2}$
&
$R_{2-d}$ & $\mathbb{Z}_2$
\\
DIII & 
$\{ i\gamma_0, i\gamma_t, T', iT'; \gamma_1, \ldots, \gamma_d, \tilde \Gamma, \tilde \Gamma \Gamma' \}$
&
$Cl_{3,d+2} \to Cl_{4,d+2}$
&
$R_{3-d}$ & $0$
\\
AII & 
$\{ i\gamma_0, i\gamma_t, T', iT'; \gamma_1, \ldots, \gamma_d, \tilde \Gamma \}$
&
$Cl_{3,d+1}\to Cl_{4,d+1}$
&
$R_{4-d}$ & $\mathbb{Z}$
\\
CII & 
$\{ i\gamma_0, i\gamma_t, \tilde \Gamma \Gamma', T', iT'; \gamma_1, \ldots, \gamma_d, \tilde \Gamma \}$
&
$Cl_{4,d+1}\to Cl_{5,d+1}$
&
$R_{5-d}$ & $0$
\\
C & 
$\{ i\gamma_0, i\gamma_t, i\tilde \Gamma, C', iC'; \gamma_1, \ldots, \gamma_d \}$
&
$Cl_{4,d}\to Cl_{5,d}$
&
$R_{6-d}$ & $0$
\\
CI & 
$\{ i\gamma_0, i\gamma_t ; T', iT' \gamma_1, \ldots, \gamma_d, \tilde \Gamma, \tilde \Gamma \Gamma' \}$ 
&
$Cl_{1,d+4}\to Cl_{2,d+4}$
&
$R_{-1-d}$ & $0$
\\
\hline \hline
\end{tabular}
\end{center}
\end{table*}

\subsection{Classification with Clifford algebras}
The above tenfold way classification of noninteracting Floquet topological phases can be derived systematically by using the Clifford algebras, which we rederive below. This will then serve as a starting point for the incorporation of the time-glide and time-screw symmetries.

We again consider the effective Hamiltonians $H_S(\bm k, t)$ in Eq.~(\ref{eq: H_S}) and topologically characterize them by considering representative Dirac Hamiltonians (which can be achieved by a suitable deformation)~\cite{Ryu10,teo-kane10,Morimoto13}.
In this approach, we study how many distinct sets of gapped Dirac Hamiltonians exist that cannot be adiabatically deformed with each other under the symmetry constraints.
Specifically, we study the Hamiltonian $H_S(\bm k, t)$ in the Dirac form written as
\begin{align}
H_S(k_1, \ldots, k_d, t)&=
k_1 \gamma_1 + \ldots + k_d \gamma_d + t \gamma_t + \gamma_0,
\end{align}
where $\gamma_i$ are Gamma matrices anticommuting with each other, in particular, $\gamma_0$ denotes the Dirac mass term.
The symmetry constraints are written as
\begin{align}
\{\gamma_i, \tilde \Gamma \}&=0,
&
[\gamma_i, \Gamma' ]&=0,
\end{align}
\begin{align}
\{\gamma_i, T' \}_{i=1,\ldots,d}&=0,
&
[\gamma_i, T']_{i=t,0}&=0,
\\
\{\gamma_i, C' \}_{i=1,\ldots,d}&=0,
&
[\gamma_i, C']_{i=t,0}&=0,
\end{align}
and 
\begin{align}
\{T', \tilde \Gamma \}=
[C', \tilde \Gamma ]=
\{\Gamma', \tilde \Gamma \}=0,
\end{align}
with
\begin{align}
\{T', i \}=
\{C', i \}=
[\Gamma', i ]&=0, 
&
[T',C']&=0.
\end{align}
These symmetry constraints are concisely described in terms of Clifford algebras in Table~\ref{table: extension tenfold way}. 
Clifford algebra is an algebra generated by generators anticommuting with each other. Specifically, complex Clifford algebra $Cl_q$ is generated over complex numbers $\mathbb{C}$ by $q$ anticommuting generators $\{e_i\}$ that satisfy
\begin{align}
\{e_i, e_j\}&= 2 \delta_{ij}.
\end{align}
These are  essentially algebras formed by gamma matrices.
Real Clifford algebra $Cl_{p,q}$ is generated over real numbers $\mathbb{R}$ by $p+q$ anticommuting generators $\{e_1,\ldots,e_p; e_{p+1},\ldots,e_{p+q}\}$ that satisfy
\begin{align}
\{e_i, e_j\}&= 0, \qquad (i \neq j) \\
e_i^2&=
\begin{cases}
-1 & (1\le i \le p) \\
+1 & (p+1\le i \le p+q) \\
\end{cases} 
\end{align}
Real Clifford algebras are considered to be algebras formed by gamma matrices in a similar way to complex Clifford algebras, but they can also accommodate symmetry operators such as $T$ and $C$ that involves complex conjugation.  

Now the Clifford algebra is used to deduce topological classification as follows. First, we fix the representation of  kinetic terms $\gamma_1,\ldots,\gamma_d$ and symmetry operators $T,C,\Gamma$. In this case, distinct gapped Dirac Hamiltonians have one-to-one correspondence to  distinct Dirac mass terms $\gamma_0$; disconnected components of the space of possible Dirac mass terms correspond to distinct topological phases \cite{Kitaev09,Morimoto13}.
This space of Dirac mass term can be obtained from ``the extension problem'' of Clifford algebras which is summarized in Table~\ref{table: extension tenfold way}.
Namely, we fix representations of the Clifford algebras $Cl_q$ and $Cl_{p,q}$ that is generated by kinetic terms and symmetry operators, and extends it by adding the generator involving the mass term $\gamma_0$. All possible extensions form a space $V$ as 
\begin{align}
Cl_{q} &\to Cl_{q+1} &  V&=C_{q}, \\
Cl_{p,q} &\to Cl_{p,q+1} &  V&=R_{q-p}, \\
Cl_{p,q} &\to Cl_{p+1,q} &  V&=R_{p-q+2},
\end{align}
where $C_q$ and $R_q$ are symmetric spaces that appear in complex and real K-theory (for details, see Ref.~\cite{Morimoto13}).
Thus the zeroth homotopy groups for spaces $C_q, R_q$ associated with extension problems give the Abelian groups that characterize topological phases. 
The relevant extension problems and spaces of Dirac masses are shown in Table~\ref{table: extension tenfold way}.
This reproduces the classification of Floquet topological phases in tenfold way as shown in Table~\ref{table: AZ classes}.

\begin{table*}[tb]
\begin{center}
\caption{\label{table: Clifford time glide}
Clifford algebras characterizing the Floquet topological phases with time glide symmetry.
Commutation/anticommutation relationships between time glide operator and generic symmetry operators are specified by $\eta_\Gamma$ for complex classes and $(\eta_T,\eta_C)$ for real classes, where the entry $0$ indicates the absence of such symmetry in the symmetry class.  Addition of time glide operator modifies the Clifford algebras as shown in the third column. (Here, $\gamma_i$ is the short hand notation for kinetic gamma matrices $\gamma_1,\ldots,\gamma_d$.) The extension problem of the Clifford algebra with respect to the mass term $\gamma_0$ gives the space of Dirac masses $V$ shown in the last column.
\\
}
\begin{tabular}[t]{ c c c c c }
\hline \hline
~Class~ 
&
~$\eta_\Gamma$ or $(\eta_T,\eta_C)$~
& 
~Clifford algebras~
& 
~Extension problem~
& 
~Space of masses $V$~
\\
\hline
A
&
0
& 
$\{ \gamma_0, \gamma_i, \gamma_t, \tilde \Gamma, \gamma_1\gamma_t\tilde\Gamma M_T' \}$
&
$Cl_{d+3} \to Cl_{d+4}$
&
$C_{d+1}$ 
\\
AIII
&
$+$
&
$\{ \gamma_0, \gamma_i, \gamma_t, \tilde \Gamma, \tilde \Gamma \Gamma' \} \tensor  \{\gamma_1\gamma_t \Gamma' M_T' \}$
&
$Cl_{d+3}\tensor Cl_{1} \to Cl_{d+4}\tensor Cl_{1}$
&
$C_{d+1}\times C_{d+1}$
\\
AIII
&
$-$
&
$\{ \gamma_0, \gamma_i, \gamma_t, \tilde \Gamma, \tilde \Gamma \Gamma', \gamma_1\gamma_t\tilde\Gamma M_T' \}$
&
$Cl_{d+4} \to Cl_{d+5}$
&
$C_{d}$
\\
\hline
AI   
&
$(+,0)$
&
$\{ i\gamma_0, i\gamma_t, i\gamma_1\gamma_t \tilde\Gamma M_T'; T', iT', \gamma_i, \tilde \Gamma \}$
&
$Cl_{2,d+3} \to Cl_{3,d+3}$
&
$R_{1-d}$ 
\\
AI   
&
$(-,0)$
&
$\{ i\gamma_0, i\gamma_t; T', iT', \gamma_i, \tilde \Gamma, i\gamma_1\gamma_t \tilde\Gamma M_T' \}$
&
$Cl_{1,d+4} \to Cl_{2,d+4}$
&
$R_{-1-d}$ 
\\
BDI 
& 
$(+,+)$ &
$\{ i\gamma_0, i\gamma_t, \tilde \Gamma \Gamma'; T', iT', \gamma_i, \tilde \Gamma \} \tensor \{ ;i\gamma_1 \gamma_t \Gamma' M_T' \}$
&
$Cl_{2,d+3}\tensor Cl_{0,1} \to Cl_{3,d+3}\tensor Cl_{0,1}$
&
$R_{1-d}\times R_{1-d}$ 
\\
BDI 
& 
$(+,-)$ &
$\{ i\gamma_0, i\gamma_t, \tilde \Gamma \Gamma',i\gamma_1 \gamma_t \tilde\Gamma M_T'; T', iT', \gamma_i, \tilde \Gamma \}$
&
$Cl_{3,d+3} \to Cl_{4,d+3}$
&
$R_{2-d}$ 
\\
BDI 
& 
$(-,+)$ &
$\{ i\gamma_0, i\gamma_t, \tilde \Gamma \Gamma'; T', iT', \gamma_i, \tilde \Gamma,\gamma_1 \gamma_t \Gamma' M_T' \}$
&
$Cl_{2,d+4} \to Cl_{3,d+4}$
&
$R_{-d}$ 
\\
BDI 
& 
$(-,-)$ &
$\{ i\gamma_0, i\gamma_t, \tilde \Gamma \Gamma'; T', iT', \gamma_i, \tilde \Gamma \} \tensor \{ \gamma_1 \gamma_t \Gamma' M_T' ; \}$
&
$Cl_{2,d+3}\tensor Cl_{1,0} \to Cl_{3,d+3}\tensor Cl_{1,0}$
&
$C_{1+d}$ 
\\
D 
&
$(0,+)$ &
$\{ i\gamma_0, i\gamma_t, i\tilde \Gamma; C', iC', \gamma_i, \gamma_1\gamma_t \tilde\Gamma M_T' \}$
&
$Cl_{2,d+3} \to Cl_{3,d+3}$
&
$R_{1-d}$
\\
D 
&
$(0,-)$ &
$\{ i\gamma_0, i\gamma_t, i\tilde \Gamma, i\gamma_1\gamma_t \tilde\Gamma M_T' ; C', iC', \gamma_i \}$
&
$Cl_{3,d+2} \to Cl_{4,d+2}$
&
$R_{3-d}$
\\
DIII 
& 
$(+,+)$ &
$\{ i\gamma_0, i\gamma_t, T', iT'; \gamma_i, \tilde \Gamma, \tilde \Gamma \Gamma' \} \tensor \{ i\gamma_1 \gamma_t \Gamma' M_T' ; \}$
&
$Cl_{3,d+2}\tensor Cl_{1,0} \to Cl_{4,d+2}\tensor Cl_{1,0}$
&
$C_{1+d}$ 
\\
DIII 
& 
$(+,-)$ &
$\{ i\gamma_0, i\gamma_t, T', iT', i\gamma_1 \gamma_t \tilde\Gamma M_T' ; \gamma_i, \tilde \Gamma, \tilde \Gamma \Gamma' \} $
&
$Cl_{4,d+2} \to Cl_{5,d+2}$
&
$R_{4-d}$ 
\\
DIII 
& 
$(-,+)$ &
$\{ i\gamma_0, i\gamma_t, T', iT' ; \gamma_i, \tilde \Gamma, \tilde \Gamma \Gamma', \gamma_1 \gamma_t \Gamma' M_T' \} $
&
$Cl_{3,d+3} \to Cl_{4,d+3}$
&
$R_{2-d}$ 
\\
DIII 
& 
$(-,-)$ &
$\{ i\gamma_0, i\gamma_t, T', iT'; \gamma_i, \tilde \Gamma, \tilde \Gamma \Gamma' \} \tensor \{ ; \gamma_1 \gamma_t \Gamma' M_T' \}$
&
$Cl_{3,d+2}\tensor Cl_{0,1} \to Cl_{4,d+2}\tensor Cl_{0,1}$
&
$R_{3-d}\times R_{3-d} $ 
\\
AII   
&
$(+,0)$
&
$\{ i\gamma_0, i\gamma_t, T', iT', i\gamma_1\gamma_t \tilde\Gamma M_T';  \gamma_i, \tilde \Gamma \}$
&
$Cl_{4,d+1} \to Cl_{5,d+1}$
&
$R_{5-d}$ 
\\
AII   
&
$(-,0)$
&
$\{ i\gamma_0, i\gamma_t, T', iT'; \gamma_i, \tilde \Gamma, i\gamma_1\gamma_t \tilde\Gamma M_T' \}$
&
$Cl_{3,d+2} \to Cl_{4,d+2}$
&
$R_{3-d}$ 
\\
CII 
& 
$(+,+)$ &
$\{ i\gamma_0, i\gamma_t,  T', iT', \tilde \Gamma \Gamma'; \gamma_i, \tilde \Gamma \} \tensor \{ ;i\gamma_1 \gamma_t \Gamma' M_T' \}$
&
$Cl_{4,d+1}\tensor Cl_{0,1} \to Cl_{5,d+1}\tensor Cl_{0,1}$
&
$R_{5-d}\times R_{5-d}$ 
\\
CII 
& 
$(+,-)$ &
$\{ i\gamma_0, i\gamma_t,  T', iT', \tilde \Gamma \Gamma',i\gamma_1 \gamma_t \tilde\Gamma M_T';\gamma_i, \tilde \Gamma \}$
&
$Cl_{5,d+1} \to Cl_{6,d+1}$
&
$R_{6-d}$ 
\\
CII 
& 
$(-,+)$ &
$\{ i\gamma_0, i\gamma_t, T', iT', \tilde \Gamma \Gamma'; \gamma_i, \tilde \Gamma,\gamma_1 \gamma_t \Gamma' M_T' \}$
&
$Cl_{4,d+2} \to Cl_{5,d+2}$
&
$R_{4-d}$ 
\\
CII 
& 
$(-,-)$ &
$\{ i\gamma_0, i\gamma_t, T', iT', \tilde \Gamma \Gamma'; \gamma_i, \tilde \Gamma \} \tensor \{ \gamma_1 \gamma_t \Gamma' M_T' ; \}$
&
$Cl_{4,d+1}\tensor Cl_{1,0} \to Cl_{5,d+1}\tensor Cl_{1,0}$
&
$C_{1+d}$ 
\\
C
&
$(0,+)$ &
$\{ i\gamma_0, i\gamma_t, C', iC', i\tilde \Gamma;  \gamma_i, \gamma_1\gamma_t \tilde\Gamma M_T' \}$
&
$Cl_{4,d+1} \to Cl_{5,d+1}$
&
$R_{5-d}$
\\
C 
&
$(0,-)$ &
$\{ i\gamma_0, i\gamma_t,  C', iC', i\tilde \Gamma, i\gamma_1\gamma_t \tilde\Gamma M_T' ;\gamma_i \}$
&
$Cl_{5,d} \to Cl_{6,d}$
&
$R_{7-d}$
\\
CI 
& 
$(+,+)$ &
$\{ i\gamma_0, i\gamma_t; \gamma_i, T', iT', \tilde \Gamma, \tilde \Gamma \Gamma' \} \tensor \{ i\gamma_1 \gamma_t \Gamma' M_T' ; \}$
&
$Cl_{1,d+4}\tensor Cl_{1,0} \to Cl_{2,d+4}\tensor Cl_{1,0}$
&
$C_{1+d}$ 
\\
CI
& 
$(+,-)$ &
$\{ i\gamma_0, i\gamma_t, i\gamma_1 \gamma_t \tilde\Gamma M_T' ; \gamma_i, T', iT', \tilde \Gamma, \tilde \Gamma \Gamma' \} $
&
$Cl_{2,d+4} \to Cl_{3,d+4}$
&
$R_{-d}$ 
\\
CI 
& 
$(-,+)$ &
$\{ i\gamma_0, i\gamma_t ; \gamma_i, T', iT', \tilde \Gamma, \tilde \Gamma \Gamma', \gamma_1 \gamma_t \Gamma' M_T' \} $
&
$Cl_{1,d+5} \to Cl_{2,d+5}$
&
$R_{-2-d}$ 
\\
CI 
& 
$(-,-)$ &
$\{ i\gamma_0, i\gamma_t; \gamma_i, T', iT', \tilde \Gamma, \tilde \Gamma \Gamma' \} \tensor \{ ; \gamma_1 \gamma_t \Gamma' M_T' \}$
&
$Cl_{1,d+4}\tensor Cl_{0,1} \to Cl_{2,d+4}\tensor Cl_{0,1}$
&
$R_{-1-d}\times R_{-1-d} $ 
\\
\hline \hline
\end{tabular}
\end{center}
\end{table*}

\begin{table*}[tb]
\begin{center}
\caption{\label{table: Classification time glide}
Classification of Floquet topological phases with time glide symmetry. The time glide operator $M_T$ is characterized by commutation/anticommutation relations with other symmetry operators as
$M_T T=\eta_T T M_T$, $M_T C=\eta_C C M_T$,
and
$M_T \Gamma=\eta_\Gamma \Gamma M_T$.
}

\footnotesize
\begin{tabular}[t]{ccccccccccc}
\hline \hline
$\eta_T, \eta_C, \eta_\Gamma$ & Class & $C_q$ or $R_q$ & $d=0$ & $d=1$ & $d=2$ & $d=3$
 & $d=4$ & $d=5$ & $d=6$ & $d=7$ \\
\hline
 -  & A      &$C_{d+3}$& $0$ & $\mathbb{Z}$ & $0$ & $\mathbb{Z}$ & $0$
 & $\mathbb{Z}$ & $0$ & $\mathbb{Z}$ \\
$\eta_\Gamma=+$& AIII   &$C_{d+3} \times C_{d+3}$& $0$ & $\mathbb{Z}^2$ & $0$ & $\mathbb{Z}^2$ & $0$ & $\mathbb{Z}^2$
 & $0$ & $\mathbb{Z}^2$  \\
$\eta_\Gamma=-$& AIII   &$C_{d+4}$& $\mathbb{Z}$ & $0$ & $\mathbb{Z}$ & $0$
 & $\mathbb{Z}$ & $0$ & $\mathbb{Z}$ & $0$ \\
\hline 
\multirow{8}{*}{
\begin{tabular}[t]{c}
$\eta_T=+$ (AI,AII) \\ $\eta_C=-$ (D,C) \\
$(\eta_T,\eta_C)=(+,-)$ (BDI,DIII,CII,CI)
\end{tabular}
}
 & AI   &$R_{1-d}$& $\mathbb{Z}_2$ & $\mathbb{Z}$ & $0$ & $0$
 & $0$ & $\mathbb{Z}$ & $0$ & $\mathbb{Z}_2$ \\
 & BDI  &$R_{2-d}$& $\mathbb{Z}_2$ & $\mathbb{Z}_2$ & $\mathbb{Z}$ & $0$
 & $0$ & $0$ & $\mathbb{Z}$ & $0$ \\
 & D    &$R_{3-d}$& $0$ & $\mathbb{Z}_2$ & $\mathbb{Z}_2$ & $\mathbb{Z}$
 & $0$ & $0$ & $0$ & $\mathbb{Z}$ \\
 & DIII &$R_{4-d}$& $\mathbb{Z}$ & $0$ & $\mathbb{Z}_2$ & $\mathbb{Z}_2$
 & $\mathbb{Z}$ & $0$ & $0$ & $0$ \\
 & AII  &$R_{5-d}$& $0$ & $\mathbb{Z}$ & $0$ & $\mathbb{Z}_2$
 & $\mathbb{Z}_2$ & $\mathbb{Z}$ & $0$ & $0$  \\
 & CII  &$R_{6-d}$& $0$ & $0$ & $\mathbb{Z}$ & $0$
 & $\mathbb{Z}_2$ & $\mathbb{Z}_2$ & $\mathbb{Z}$ & $0$ \\
 & C    &$R_{7-d}$& $0$ & $0$  & $0$ & $\mathbb{Z}$ & $0$
 & $\mathbb{Z}_2$ & $\mathbb{Z}_2$ & $\mathbb{Z}$ \\
 & CI   &$R_{-d}$& $\mathbb{Z}$ & $0$ & $0$  & $0$ &
 $\mathbb{Z}$ & $0$ & $\mathbb{Z}_2$ & $\mathbb{Z}_2$ \\
\hline
\multirow{8}{*}{
\begin{tabular}[t]{c}
$\eta_T=-$ (AI,AII) \\ $\eta_C=+$ (D,C) \\
$(\eta_T,\eta_C)=(-,+)$ (BDI,DIII,CII,CI)
\end{tabular}
}
 & AI   &$R_{-1-d}$& $0$ & $0$  & $0$ & $\mathbb{Z}$
 & $0$ & $\mathbb{Z}_2$ & $\mathbb{Z}_2$ & $\mathbb{Z}$ \\
 & BDI  &$R_{-d}$& $\mathbb{Z}$ & $0$ & $0$  & $0$
 & $\mathbb{Z}$ & $0$ & $\mathbb{Z}_2$ & $\mathbb{Z}_2$ \\
 & D    &$R_{1-d}$& $\mathbb{Z}_2$ & $\mathbb{Z}$ & $0$
 & $0$ & $0$ & $\mathbb{Z}$ & $0$ & $\mathbb{Z}_2$ \\
 & DIII &$R_{2-d}$& $\mathbb{Z}_2$ & $\mathbb{Z}_2$ & $\mathbb{Z}$ & $0$
 & $0$ & $0$ & $\mathbb{Z}$ & $0$ \\
 & AII  &$R_{3-d}$& $0$ & $\mathbb{Z}_2$ & $\mathbb{Z}_2$ & $\mathbb{Z}$
 & $0$ & $0$ & $0$ & $\mathbb{Z}$ \\
 & CII  &$R_{4-d}$& $\mathbb{Z}$ & $0$ & $\mathbb{Z}_2$ & $\mathbb{Z}_2$
 & $\mathbb{Z}$ & $0$ & $0$ & $0$ \\
 & C    &$R_{5-d}$& $0$ & $\mathbb{Z}$ & $0$ & $\mathbb{Z}_2$
 & $\mathbb{Z}_2$ & $\mathbb{Z}$ & $0$ & $0$  \\
 & CI   &$R_{6-d}$& $0$ & $0$ & $\mathbb{Z}$ & $0$
 & $\mathbb{Z}_2$ & $\mathbb{Z}_2$ & $\mathbb{Z}$ & $0$ \\
\hline
$(\eta_T,\eta_C)=(+,+)$ & BDI  &$R_{1-d} \times R_{1-d}$& $\mathbb{Z}_2^2$ & $\mathbb{Z}^2$ & $0$ & $0$
 & $0$ & $\mathbb{Z}^2$ & $0$ & $\mathbb{Z}_2^2$ \\
$(\eta_T,\eta_C)=(-,-)$ & DIII &$R_{3-d} \times R_{3-d}$& $0$ & $\mathbb{Z}_2^2$ & $\mathbb{Z}_2^2$ & $\mathbb{Z}^2$
 & $0$ & $0$ & $0$ & $\mathbb{Z}^2$ \\
$(\eta_T,\eta_C)=(+,+)$ & CII  &$R_{5-d} \times R_{5-d}$& $0$ & $\mathbb{Z}^2$ & $0$ & $\mathbb{Z}_2^2$
 & $\mathbb{Z}_2^2$ & $\mathbb{Z}^2$ & $0$ & $0$  \\
$(\eta_T,\eta_C)=(-,-)$ & CI   &$R_{7-d} \times R_{7-d}$& $0$ & $0$  & $0$ & $\mathbb{Z}^2$
 & $0$ & $\mathbb{Z}_2^2$ & $\mathbb{Z}_2^2$ & $\mathbb{Z}^2$ \\
\hline
$(\eta_T,\eta_C)=(-,-)$ & BDI, CII  &$C_{1-d}\times C_{1-d}$& $0$ & $\mathbb{Z}^2$ & $0$ & $\mathbb{Z}^2$
 & $0$ & $\mathbb{Z}^2$ & $0$ & $\mathbb{Z}^2$ \\
$(\eta_T,\eta_C)=(+,+)$ & DIII, CI  &$C_{1-d}\times C_{1-d}$& $0$ & $\mathbb{Z}^2$ & $0$ & $\mathbb{Z}^2$
 & $0$ & $\mathbb{Z}^2$ & $0$ & $\mathbb{Z}^2$ \\
\hline \hline
\end{tabular}
\label{tab: time-glide}
\end{center}
\end{table*}

\subsection{Classification of time glide symmetric Floquet topological phases \label{sec: classification time glide}}
In this section, we classify Floquet topological phases with time glide symmetry by using Clifford algebra approach.
Recall the time glide in the $x_1$-direction is represented by a unitary operator $M_T$ satisfying
\begin{align}
M_T H(k_1,k_2,\ldots, k_d, t) M_T^{-1}
&= H \left(-k_1,k_2,\ldots, k_d, t+\frac{T}{2} \right).
\end{align}
We assume $M_T^2=1$ without loss of generality.
This constrains the symmetrized time-evolution operator $U(\bm k,t)$ as
\begin{align}
M_T U(\bm k, t) M_T^{-1}
&=
M_T 
\mathcal{T}\exp \left[-i\int_{\frac{T-t}{2}}^{\frac{T+t}{2}} dt' H(\bm k, t') \right]
M_T^{-1}
\n
&=
\mathcal{T}\exp \left[-i\int_{\frac{-t}{2}}^{\frac{+t}{2}} dt' H(\hat R \bm k, t') \right]
\n
&=
\left(
\mathcal{T}\exp \left[-i\int_{\frac{+t}{2}}^{\frac{2T-t}{2}} dt' H(\hat R \bm k, t') \right]
\right)^\dagger
\n
&= U^\dagger(\hat R \bm k, T-t) 
\end{align}
with $\hat R(k_1,k_2,\ldots, k_d)=(-k_1,k_2,\ldots, k_d)$,
where we used 
$\mathcal{T}\exp \left[-i\int_{\frac{-t}{2}}^{\frac{2T-t}{2}} dt' H(\hat R \bm k, t') \right]=1$ (triviality of the full period evolution).
Thus we have an additional symmetry constraint onto the effective Hamiltonian $H_S(\bm k, t)$ given by
\begin{align}
M_T' H_S(k_1, k_2, \ldots,k_d,t) M_T'^{-1} &= H_S(-k_1, k_2, \ldots,k_d, -t), \n
M_T'&=M_T \tensor \sigma_x.
\label{eq: M_T' and H}
\end{align}

Inclusion of the time glide symmetry modifies the Clifford algebras characterizing Floquet topological phases as summarized in Table~\ref{table: Clifford time glide}.
First, Eq.~(65) indicates commutation/anticommutation relationships between the time glide operator $M_T'$ and gamma matrices as
\begin{align}
[\gamma_0,M_T']=[\gamma_i,M_T']=0
\end{align}
for $i=2,\ldots,d$ and
\begin{align}
\{\gamma_1,M_T'\}=\{\gamma_t,M_T'\}=0.
\end{align}
In addition, $M_T'$ anticommutes with the intrinsic chiral operator $\tilde \Gamma=\sigma_z$.
Therefore the operator $\gamma_1\gamma_t\tilde \Gamma M_T'$ anticommutes with the gamma matrices and $\tilde \Gamma$ and gives a candidate of an additional generator in the Clifford algebras. Indeed, in the case of class A, this is the additional generator.
Namely, in class A, the time glide symmetry reserves one additional gamma matrix for the operator $\gamma_1\gamma_t\tilde \Gamma M_T'$ which is equivalent to raising the spatial dimension by one. Thus time-glide symmetric Floquet topological phases in class A possess integer topological numbers in odd dimensions, and are trivial otherwise.
Next, the form of the additional generator including $M_T$ in the Clifford algebras depends on commutation/anticommutation relation between the time glide operator $M_T'$ and other symmetry operators if present.
Specifically, in the case of class AIII, the relationship between $\Gamma$ and $M_T$ is given by
\begin{align}
M_T \Gamma &= \eta_\Gamma \Gamma M_T, 
\end{align}
where $\eta_\Gamma=\pm 1$ denotes commutation and anticommutation relations, respectively.
In the case of $\eta_\Gamma=+1$, the operator $\gamma_1\gamma_t \Gamma' M_T'$ commute with every original generator. This indicates that the energy eigenstates of the effective Dirac Hamiltonian are simultaneously eigenstates of $\gamma_1\gamma_t \Gamma' M_T'$, and hence, are decomposed into two sectors with $\gamma_1\gamma_t \Gamma' M_T'=\pm 1$. Thus, time-glide symmetric Floquet topological phases in class AIII with $\eta_\Gamma=+1$ are characterized by two topological numbers of class AIII for each sector.
In the case of $\eta_\Gamma=-1$, the operator $\gamma_1\gamma_t\tilde \Gamma M_T'$ anticommute with every original generator. Thus the time glide symmetry requires one additional gamma matrix for $\gamma_1\gamma_t\tilde \Gamma M_T'$ which is equivalent to raising the spatial dimension by one for original Floquet topological phases in class AIII. Namely, time-glide symmetric Floquet topological phases in class AIII with $\eta_\Gamma=-1$ possess integer and trivial topological numbers in even and odd spatial dimensions, respectively.
We note that this classification theory with Clifford algebras is indeed consistent with discussions in Sec.~\ref{sec: models}; the Clifford algebra analysis gives integer topological numbers for 2D class AIII ($\eta_\Gamma=-1$) and 3D class A systems with time glide symmetry as expected from explicit constructions of integer topological numbers presented in Sec.~\ref{sec: models}.

In a similar way, we can classify Floquet topological phases with time glide symmetry in the real classes.
The forms of relevant Clifford algebras are determined by commutation/anticommutation relationships between $T,C$ and $M_T$ described by $(\eta_T,\eta_C)$ as
\begin{align}
M_T T &= \eta_T T M_T, &
M_C C &= \eta_C C M_C.
\end{align}
In this case, depending on the combination $(\eta_T,\eta_C)$, either one of $\gamma_1\gamma_t \tilde\Gamma M_T'$, $i\gamma_1\gamma_t \tilde\Gamma M_T'$, $\gamma_1 \gamma_t \Gamma' M_T'$, $i\gamma_1 \gamma_t \Gamma' M_T'$ commute or anticommute with all the original generators and gives the additional generator of the Clifford algebras. 
The explicit forms of the Clifford algebras are summarized in Table~\ref{table: Clifford time glide}.  

Finally, we present the result of classification of Floquet topological phases with time glide symmetry in Table~\ref{table: Classification time glide}. This is obtained by taking zeroth homotopy group of the space of Dirac masses $V$ shown in Table~\ref{table: Clifford time glide}. The classification table shows a periodic structure with respect to the spatial dimensions. There are four types of changes from original classification table for ten fold way. i.e., (i) shift of the spatial dimensions by $+1$, (ii) shift of the spatial dimension by $-1$,
(iii) doubling of topological numbers due to the block diagonalization, and (iv) reduction of real symmetry classes to complex symmetry classes due to the block diagonalization.
These phenomena are also found in classification theory of topological crystalline insulators (TCIs) in the equilibrium, and indeed the obtained classification table for time glide symmetry resembles that for TCIs~\cite{Chiu13,Morimoto13,Shiozaki14}. However, the classification result for time glide symmetric Floquet topological phases is different from that for TCIs. For example, if we look at class AIII, the doubling of topological number and the shift of the spatial dimension take place for $\eta_\Gamma=+1$ and $\eta_\Gamma=-1$, respectively, which are opposite for the classification of TCIs. 
This suggests that the time glide symmetric Floquet topological phases are intrinsically nonequilibrium topological phases and are not adiabatically connected to equilibrium topological phases.



\begin{table*}[tb]
\begin{center}
\caption{\label{table: Clifford time screw}
Clifford algebras characterizing the Floquet topological phases with time screw symmetry.
Commutation/anticommutation relationships between time screw operator and generic symmetry operators are specified by $\eta_\Gamma$ for complex classes and $(\eta_T,\eta_C)$ for real classes, where the entry $0$ indicates the absence of such symmetry in the symmetry class.  Addition of time screw operator modifies the Clifford algebras as shown in the third column. (Here, $\gamma_i$ is the short hand notation for kinetic gamma matrices $\gamma_1,\ldots,\gamma_d$.) The extension problem of the Clifford algebra with respect to the mass term $\gamma_0$ gives the space of Dirac masses $V$ shown in the last column.
\\
}
\begin{tabular}[t]{ c c c c c }
\hline \hline
~Class~ 
&
~$\eta_\Gamma$ or $(\eta_T,\eta_C)$~
& 
~Clifford algebras~
& 
~Extension problem~
& 
~Space of masses $V$~
\\
\hline
A
&
0
& 
$\{ \gamma_0, \gamma_i, \gamma_t, \tilde \Gamma \}\tensor \{ \gamma_1\gamma_2\gamma_t\tilde\Gamma C_{2T}' \}$
&
$Cl_{d+2}\tensor Cl_{1} \to Cl_{d+3}\tensor Cl_{1}$
&
$C_{d}\times C_{d}$ 
\\
AIII
&
$+$
&
$\{ \gamma_0, \gamma_i, \gamma_t, \tilde \Gamma, \tilde \Gamma \Gamma' , \gamma_1 \gamma_2 \gamma_t \Gamma' C_{2T}' \}$
&
$Cl_{d+4} \to Cl_{d+5}$
&
$C_{d}$
\\
AIII
&
$-$
&
$\{ \gamma_0, \gamma_i, \gamma_t, \tilde \Gamma, \tilde \Gamma \Gamma' \}\tensor\{ \gamma_1 \gamma_2 \gamma_t \tilde \Gamma C_{2T}' \}$
&
$Cl_{d+3} \tensor Cl_{1} \to Cl_{d+4} \tensor Cl_{1}$
&
$C_{d+1} \times C_{d+1}$
\\
\hline
AI   
&
$(+,0)$
&
$\{ i\gamma_0, i\gamma_t; T', iT', \gamma_i, \tilde \Gamma \} \tensor\{ i\gamma_1 \gamma_2 \gamma_t \tilde \Gamma C_{2T}'; \}$
&
$Cl_{1,d+3}\tensor Cl_{1,0} \to Cl_{2,d+3}\tensor Cl_{1,0}$
&
$C_{d}$ 
\\
AI   
&
$(-,0)$
&
$\{ i\gamma_0, i\gamma_t; T', iT', \gamma_i, \tilde \Gamma \} \tensor\{ ; \gamma_1 \gamma_2 \gamma_t \tilde \Gamma C_{2T}' \}$
&
$Cl_{1,d+3}\tensor Cl_{0,1} \to Cl_{2,d+3}\tensor Cl_{0,1}$
&
$R_{-d}\times R_{-d}$ 
\\
BDI 
& 
$(+,+)$ &
$\{ i\gamma_0, i\gamma_t, \tilde \Gamma \Gamma', i\gamma_1 \gamma_2 \gamma_t \Gamma' C_{2T}' ; T', iT', \gamma_i, \tilde \Gamma \}$
&
$Cl_{3,d+3} \to Cl_{4,d+3}$
&
$R_{2-d}$ 
\\
BDI 
& 
$(+,-)$ &
$\{ i\gamma_0, i\gamma_t, \tilde \Gamma \Gamma'; T', iT', \gamma_i, \tilde \Gamma \} \tensor \{ i\gamma_1 \gamma_2 \gamma_t \tilde \Gamma C_{2T}' ; \}$
&
$Cl_{2,d+3}\tensor Cl_{1,0} \to Cl_{3,d+3}\tensor Cl_{1,0}$
&
$C_{1+d}$ 
\\
BDI 
& 
$(-,+)$ &
$\{ i\gamma_0, i\gamma_t, \tilde \Gamma \Gamma'; T', iT', \gamma_i, \tilde \Gamma \} \tensor \{ \gamma_1 \gamma_2 \gamma_t \tilde \Gamma C_{2T}' ; \}$
&
$Cl_{2,d+3}\tensor Cl_{0,1} \to Cl_{3,d+3}\tensor Cl_{0,1}$
&
$R_{1-d}\times R_{1-d}$ 
\\
BDI 
& 
$(-,-)$ &
$\{ i\gamma_0, i\gamma_t, \tilde \Gamma \Gamma' ; T', iT', \gamma_i, \tilde \Gamma,  \gamma_1 \gamma_2 \gamma_t \Gamma' C_{2T}' \}$
&
$Cl_{2,d+4} \to Cl_{3,d+4}$
&
$R_{-d}$ 
\\
D 
&
$(0,+)$ &
$\{ i\gamma_0, i\gamma_t, i\tilde \Gamma; C', iC', \gamma_i\} \tensor \{; \gamma_1\gamma_2 \gamma_t \tilde\Gamma C_{2T}' \}$
&
$Cl_{2,d+2}\tensor Cl_{0,1} \to Cl_{3,d+2}\tensor Cl_{0,1}$
&
$R_{2-d}\times R_{2-d}$
\\
D 
&
$(0,-)$ &
$\{ i\gamma_0, i\gamma_t, i\tilde \Gamma; C', iC', \gamma_i\} \tensor \{ i\gamma_1\gamma_2 \gamma_t \tilde\Gamma C_{2T}' ; \}$
&
$Cl_{2,d+2}\tensor Cl_{1,0} \to Cl_{3,d+2}\tensor Cl_{1,0}$
&
$C_{d}$
\\
DIII 
& 
$(+,+)$ &
$\{ i\gamma_0, i\gamma_t, T', iT' ;  \gamma_i, \tilde \Gamma, \tilde \Gamma \Gamma', i\gamma_1 \gamma_2 \gamma_t \Gamma' C_{2T}' \}$
&
$Cl_{3,d+3} \to Cl_{4,d+3}$
&
$R_{2-d}$ 
\\
DIII
& 
$(+,-)$ &
$\{ i\gamma_0, i\gamma_t,  T', iT'; \gamma_i, \tilde \Gamma, \tilde \Gamma \Gamma' \} \tensor \{ i\gamma_1 \gamma_2 \gamma_t \tilde \Gamma C_{2T}' ; \}$
&
$Cl_{3,d+2}\tensor Cl_{1,0} \to Cl_{4,d+2}\tensor Cl_{1,0}$
&
$C_{1+d}$ 
\\
DIII
& 
$(-,+)$ &
$\{ i\gamma_0, i\gamma_t,  T', iT'; \gamma_i, \tilde \Gamma, \tilde \Gamma \Gamma' \} \tensor \{ \gamma_1 \gamma_2 \gamma_t \tilde \Gamma C_{2T}' ; \}$
&
$Cl_{3,d+2}\tensor Cl_{0,1} \to Cl_{4,d+2}\tensor Cl_{0,1}$
&
$R_{3-d}\times R_{3-d}$ 
\\
DIII 
& 
$(-,-)$ &
$\{ i\gamma_0, i\gamma_t, T', iT', \gamma_1 \gamma_2 \gamma_t \Gamma' C_{2T}' ;  \gamma_i, \tilde \Gamma, \tilde \Gamma \Gamma' \}$
&
$Cl_{4,d+2} \to Cl_{5,d+2}$
&
$R_{4-d}$ 
\\
AII   
&
$(+,0)$
&
$\{ i\gamma_0, i\gamma_t,  T', iT'; \gamma_i, \tilde \Gamma \} \tensor\{ i\gamma_1 \gamma_2 \gamma_t \tilde \Gamma C_{2T}'; \}$
&
$Cl_{3,d+1}\tensor Cl_{1,0} \to Cl_{4,d+1}\tensor Cl_{1,0}$
&
$C_{d}$ 
\\
AII   
&
$(-,0)$
&
$\{ i\gamma_0, i\gamma_t,  T', iT'; \gamma_i, \tilde \Gamma \} \tensor\{ ; \gamma_1 \gamma_2 \gamma_t \tilde \Gamma C_{2T}' \}$
&
$Cl_{3,d+1}\tensor Cl_{0,1} \to Cl_{4,d+1}\tensor Cl_{0,1}$
&
$R_{4-d}\times R_{4-d}$ 
\\
CII 
& 
$(+,+)$ &
$\{ i\gamma_0, i\gamma_t, T', iT', \tilde \Gamma \Gamma', i\gamma_1 \gamma_2 \gamma_t \Gamma' C_{2T}' ; \gamma_i, \tilde \Gamma \}$
&
$Cl_{5,d+1} \to Cl_{6,d+1}$
&
$R_{6-d}$ 
\\
CII
& 
$(+,-)$ &
$\{ i\gamma_0, i\gamma_t, T', iT', \tilde \Gamma \Gamma'; \gamma_i, \tilde \Gamma \} \tensor \{ i\gamma_1 \gamma_2 \gamma_t \tilde \Gamma C_{2T}' ; \}$
&
$Cl_{4,d+1}\tensor Cl_{1,0} \to Cl_{5,d+1}\tensor Cl_{1,0}$
&
$C_{1+d}$ 
\\
CII 
& 
$(-,+)$ &
$\{ i\gamma_0, i\gamma_t, T', iT', \tilde \Gamma \Gamma';  \gamma_i, \tilde \Gamma \} \tensor \{ \gamma_1 \gamma_2 \gamma_t \tilde \Gamma C_{2T}' ; \}$
&
$Cl_{4,d+1}\tensor Cl_{0,1} \to Cl_{5,d+1}\tensor Cl_{0,1}$
&
$R_{5-d}\times R_{5-d}$ 
\\
CII
& 
$(-,-)$ &
$\{ i\gamma_0, i\gamma_t, T', iT', \tilde \Gamma \Gamma' ; \gamma_i, \tilde \Gamma,  \gamma_1 \gamma_2 \gamma_t \Gamma' C_{2T}' \}$
&
$Cl_{4,d+2} \to Cl_{5,d+2}$
&
$R_{4-d}$ 
\\
C
&
$(0,+)$ &
$\{ i\gamma_0, i\gamma_t, C', iC', i\tilde \Gamma; \gamma_i\} \tensor \{; \gamma_1\gamma_2 \gamma_t \tilde\Gamma C_{2T}' \}$
&
$Cl_{4,d}\tensor Cl_{0,1} \to Cl_{5,d}\tensor Cl_{0,1}$
&
$R_{6-d}\times R_{6-d}$
\\
C 
&
$(0,-)$ &
$\{ i\gamma_0, i\gamma_t, C', iC', i\tilde \Gamma; \gamma_i\} \tensor \{ i\gamma_1\gamma_2 \gamma_t \tilde\Gamma C_{2T}' ; \}$
&
$Cl_{4,d}\tensor Cl_{1,0} \to Cl_{5,d}\tensor Cl_{1,0}$
&
$C_{d}$
\\
CI 
& 
$(+,+)$ &
$\{ i\gamma_0, i\gamma_t ;  \gamma_i, T', iT', \tilde \Gamma, \tilde \Gamma \Gamma', i\gamma_1 \gamma_2 \gamma_t \Gamma' C_{2T}' \}$
&
$Cl_{1,d+5} \to Cl_{2,d+5}$
&
$R_{-2-d}$ 
\\
CI
& 
$(+,-)$ &
$\{ i\gamma_0, i\gamma_t ; \gamma_i,  T', iT', \tilde \Gamma, \tilde \Gamma \Gamma' \} \tensor \{ i\gamma_1 \gamma_2 \gamma_t \tilde \Gamma C_{2T}' ; \}$
&
$Cl_{1,d+4}\tensor Cl_{1,0} \to Cl_{2,d+4}\tensor Cl_{1,0}$
&
$C_{1+d}$ 
\\
CI
& 
$(-,+)$ &
$\{ i\gamma_0, i\gamma_t ; \gamma_i,  T', iT', \tilde \Gamma, \tilde \Gamma \Gamma' \} \tensor \{ \gamma_1 \gamma_2 \gamma_t \tilde \Gamma C_{2T}' ; \}$
&
$Cl_{1,d+4}\tensor Cl_{0,1} \to Cl_{2,d+4}\tensor Cl_{0,1}$
&
$R_{-1-d}\times R_{-1-d}$ 
\\
CI 
& 
$(-,-)$ &
$\{ i\gamma_0, i\gamma_t, \gamma_1 \gamma_2 \gamma_t \Gamma' C_{2T}' ;  \gamma_i, T', iT', \tilde \Gamma, \tilde \Gamma \Gamma' \}$
&
$Cl_{2,d+4} \to Cl_{3,d+4}$
&
$R_{-d}$ 
\\
\hline \hline
\end{tabular}
\end{center}
\end{table*}

\begin{table*}[tb]
\begin{center}
\caption{
Classification of Floquet topological phases with time screw symmetry. The time screw operator $M_T$ is characterized by commutation/anticommutation relations with other symmetry operators as
$C_{2T} T=\eta_T T C_{2T}$, $C_{2T} C=\eta_C C C_{2T}$,
and
$C_{2T} \Gamma=\eta_\Gamma \Gamma C_{2T}$.
}

\footnotesize
\begin{tabular}[t]{@{\hspace{-3em}}ccccccccccc}
\hline \hline
$\eta_T, \eta_C, \eta_\Gamma$ & Class & $C_q$ or $R_q$ & $d=0$ & $d=1$ & $d=2$ & $d=3$
 & $d=4$ & $d=5$ & $d=6$ & $d=7$ \\
\hline
 -  & A      &$C_{d+2}\times C_{d+2}$& $\mathbb{Z}^2$ & $0$ & $\mathbb{Z}^2$ & $0$
 & $\mathbb{Z}^2$ & $0$ & $\mathbb{Z}^2$& $0$  \\
$\eta_\Gamma=+$& AIII  
&$C_{d+4}$& $\mathbb{Z}$ & $0$ & $\mathbb{Z}$ & $0$
 & $\mathbb{Z}$ & $0$ & $\mathbb{Z}$ & $0$ \\
$\eta_\Gamma=-$& AIII   
&$C_{d+3} \times C_{d+3}$& $0$ & $\mathbb{Z}^2$ & $0$ & $\mathbb{Z}^2$ & $0$ & $\mathbb{Z}^2$
 & $0$ & $\mathbb{Z}^2$  \\
\hline 
\multirow{2}{*}{
\begin{tabular}[t]{c}
$\eta_T=+$ (AI,AII), $\eta_C=-$ (D,C) \\
$(\eta_T,\eta_C)=(+,-)$ (BDI,DIII,CII,CI)
\end{tabular}
}
 & AI, AII, D, C   &$C_{d}\times C_{d}$& $\mathbb{Z}^2$ & $0$ & $\mathbb{Z}^2$ & $0$
 & $\mathbb{Z}^2$ & $0$ & $\mathbb{Z}^2$& $0$  \\
 & BDI, DIII, CII, CI  &$C_{d+1} \times C_{d+1}$& $0$ & $\mathbb{Z}^2$ & $0$ & $\mathbb{Z}^2$ & $0$ & $\mathbb{Z}^2$
 & $0$ & $\mathbb{Z}^2$  \\
\hline
\multirow{8}{*}{
\begin{tabular}[t]{c}
$\eta_T=-$ (AI,AII) \\ $\eta_C=+$ (D,C) \\
$(\eta_T,\eta_C)=(-,+)$ (BDI,DIII,CII,CI)
\end{tabular}
}
 & AI   &$R_{-d}\times R_{-d}$& $\mathbb{Z}^2$ & $0$ & $0$  & $0$
 & $\mathbb{Z}^2$ & $0$ & $\mathbb{Z}_2^2$ & $\mathbb{Z}_2^2$ \\
 & BDI  &$R_{1-d}\times R_{1-d}$& $\mathbb{Z}_2^2$ & $\mathbb{Z}^2$ & $0$
 & $0$ & $0$ & $\mathbb{Z}^2$ & $0$ & $\mathbb{Z}_2^2$ \\
 & D    &$R_{2-d}\times R_{2-d}$& $\mathbb{Z}_2^2$ & $\mathbb{Z}_2^2$ & $\mathbb{Z}^2$ & $0$
 & $0$ & $0$ & $\mathbb{Z}^2$ & $0$ \\
 & DIII &$R_{3-d}\times R_{3-d}$& $0$ & $\mathbb{Z}_2^2$ & $\mathbb{Z}_2^2$ & $\mathbb{Z}^2$
 & $0$ & $0$ & $0$ & $\mathbb{Z}^2$ \\
 & AII  &$R_{4-d}\times R_{4-d}$& $\mathbb{Z}^2$ & $0$ & $\mathbb{Z}_2^2$ & $\mathbb{Z}_2^2$
 & $\mathbb{Z}^2$ & $0$ & $0$ & $0$ \\
 & CII  &$R_{5-d}\times R_{5-d}$& $0$ & $\mathbb{Z}^2$ & $0$ & $\mathbb{Z}_2^2$
 & $\mathbb{Z}_2^2$ & $\mathbb{Z}^2$ & $0$ & $0$  \\
 & C    &$R_{6-d}\times R_{6-d}$& $0$ & $0$ & $\mathbb{Z}^2$ & $0$
 & $\mathbb{Z}_2^2$ & $\mathbb{Z}_2^2$ & $\mathbb{Z}^2$ & $0$ \\
 & CI   &$R_{7-d}\times R_{7-d}$& $0$ & $0$  & $0$ & $\mathbb{Z}^2$
 & $0$ & $\mathbb{Z}_2^2$ & $\mathbb{Z}_2^2$ & $\mathbb{Z}^2$ \\
\hline
$(\eta_T,\eta_C)=(+,+)$ & BDI  &$R_{2-d}$& $\mathbb{Z}_2$& $\mathbb{Z}_2$ & $\mathbb{Z}$ & $0$ & $0$
 & $0$ & $\mathbb{Z}$ & $0$  \\
$(\eta_T,\eta_C)=(-,-)$ & DIII &$R_{4-d}$& $\mathbb{Z}$ & $0$& $\mathbb{Z}_2$& $\mathbb{Z}_2$ & $\mathbb{Z}$ & $0$ & $0$
 & $0$   \\
$(\eta_T,\eta_C)=(+,+)$ & CII  &$R_{6-d}$& $0$
 & $0$ & $\mathbb{Z}$ & $0$& $\mathbb{Z}_2$& $\mathbb{Z}_2$ & $\mathbb{Z}$ & $0$ \\
$(\eta_T,\eta_C)=(-,-)$ & CI   &$R_{-d}$& $\mathbb{Z}$ & $0$& $0$
 & $0$ & $\mathbb{Z}$ & $0$& $\mathbb{Z}_2$& $\mathbb{Z}_2$  \\
\hline
$(\eta_T,\eta_C)=(-,-)$ & BDI  &$R_{-d}$& $\mathbb{Z}$ & $0$& $0$
 & $0$ & $\mathbb{Z}$ & $0$& $\mathbb{Z}_2$& $\mathbb{Z}_2$  \\
$(\eta_T,\eta_C)=(+,+)$ & DIII &$R_{2-d}$& $\mathbb{Z}_2$& $\mathbb{Z}_2$ & $\mathbb{Z}$ & $0$ & $0$
 & $0$ & $\mathbb{Z}$ & $0$  \\
$(\eta_T,\eta_C)=(-,-)$ & CII  &$R_{4-d}$& $\mathbb{Z}$ & $0$& $\mathbb{Z}_2$& $\mathbb{Z}_2$ & $\mathbb{Z}$ & $0$ & $0$
 & $0$   \\
$(\eta_T,\eta_C)=(+,+)$ & CI   &$R_{6-d}$& $0$
 & $0$ & $\mathbb{Z}$ & $0$& $\mathbb{Z}_2$& $\mathbb{Z}_2$ & $\mathbb{Z}$ & $0$ \\
\hline \hline
\end{tabular}
\label{tab: time-screw}
\end{center}
\end{table*}

\subsection{Classification of time screw symmetric Floquet topological phases}
In a similar manner to the time glide symmetry,
we can define ``time screw symmetry'' which is a combination of the $C_2$ rotation and the half period time translation.
In this section, we classify time screw symmetric Floquet topological phases by using Clifford algebras. The classification is performed in a similar way to that for time glide symmetry.

The time screw symmetry in the $(x_1,x_2)$ plane is represented by a unitary operator $C_{2T}$ that satisfies
\begin{align}
&C_{2T} H(k_1,k_2,k_3,\ldots, k_d, t) C_{2T}^{-1} \n
&= H \left(-k_1,-k_2,k_3,\ldots, k_d, t+\frac{T}{2} \right).
\end{align}
Accordingly, the time screw symmetry constrains the symmetrized time-evolution operator $U(\bm k,t)$ as
\begin{align}
&C_{2T} U(k_1,k_2,k_3,\ldots, k_d, t) C_{2T}^{-1} \n
&=
U^\dagger(-k_1,-k_2,k_3,\ldots, k_d, T-t),
\end{align}
and hence, the effective Hamiltonian $H_S(\bm k, t)$ made of the time evolution operator in Eq.~(\ref{eq: H_S}) is constrained as
\begin{align}
&C'_{2T} H_S(k_1, k_2,k_3, \ldots,k_d,t) (C'_{2T})^{-1} 
\n &= H_S(-k_1, -k_2, k_3, \ldots,k_d, -t),
\label{eq: C'T and H}
\end{align}
with
\begin{align}
C'_{2T}&=C_{2T} \tensor \sigma_x.
\end{align}

We obtain Clifford algebras associated with time screw symmetric Floquet topological phases by assuming that the effective Hamiltonian $H_S$ has a Dirac form.
In this case, Eq.~(\ref{eq: C'T and H}) gives commutation/anticommutation relationships between the time screw operator and gamma matrices as
\begin{align}
[\gamma_0,C_{2T}']=[\gamma_i,C_{2T}']=0
\end{align}
for $i=3,\ldots,d$, and
\begin{align}
\{\gamma_1,C_{2T}'\}=\{\gamma_2,C_{2T}'\}=\{\gamma_t,C_{2T}'\}=0.
\end{align}
In addition, $C_{2T}'$ anticommutes with the intrinsic chiral operator $\tilde \Gamma=\sigma_z$.
First, the Clifford algebras for complex symmetry classes are obtained as follows.
In the symmetry class A, the operator $\gamma_1 \gamma_2 \gamma_t \tilde \Gamma C_{2T}'$ commutes with every original generators of the Clifford algebra. Thus time screw symmetry doubles the topological number of Floquet topological phases in class A.
In the symmetry class AIII, we have two cases
\begin{align}
C_{2T}\Gamma &=\eta_\Gamma \Gamma C_{2T},
\end{align}
with $\eta_\Gamma=\pm 1$.
In the case of $\eta_\Gamma=+1$, the operator $\gamma_1 \gamma_2 \gamma_t \Gamma' C_{2T}'$ is the additional generator that anticommutes with every generators and effectively shifts the dimension by 1.
In the case of $\eta_\Gamma=-1$, the operator $\gamma_1 \gamma_2 \gamma_t \tilde \Gamma C_{2T}'$ is the additional generator that commutes with every generators and doubles the topological number due to the block diagonalization with respect to  $\gamma_1 \gamma_2 \gamma_t \tilde \Gamma C_{2T}'$.
Second, the real symmetry classes are characterized by time reversal symmetry and particle hole symmetry, and their relationships with $C_{2T}$ govern time screw symmetric Floquet topological phases. The commutation/anticommutation relationships between $C_{2T}$ and generic symmetry operators $T,C$ are specified by $(\eta_T,\eta_C)$ as
\begin{align}
C_{2T} T &=\eta_T T C_{2T}, &
C_{2T} C &=\eta_C C C_{2T}. 
\end{align}
The additional generator for the Clifford algebra that is made from the time screw operator is given by
either one of $\gamma_1\gamma_2\gamma_t \tilde\Gamma C_{2T}'$, $i\gamma_1\gamma_2\gamma_t \tilde\Gamma C_{2T}'$, $\gamma_1\gamma_2 \gamma_t \Gamma' C_{2T}'$, and $i\gamma_1\gamma_2 \gamma_t \Gamma' C_{2T}'$
which commutes or anticommutes with all the original generators.
The explicit forms of the Clifford algebras associated with time screw symmetric Floquet topological phases are listed in Table~\ref{table: Clifford time screw}. 

Finally, the classification of Floquet topological phases with time screw symmetry is summarized in Table \ref{tab: time-screw}, which is obtained by taking zeroth homotopy groups of the space of Dirac masses $V$ shown in Table~\ref{table: Clifford time screw}. 
The classification table again shows a periodic structure with respect to the spatial dimension $d$. Real symmetry classes having either $T$ or $C$ (classes AI,D,AII,C) show either reduction to complex classes or doubling of topological numbers due to the block diagonalization with respect to the additional generator involving $C_{2T}$.
Real symmetry classes having both $T$ and $C$ (classes BDI, DIII, CII, CI) exhibit four types of changes from the original ten fold way classification in a similar manner to the case of time glide symmetry.

\section{Discussions \label{sec: discussions}}

We have shown that the time glide symmetry, which is an intrinsically dynamical symmetry, can host a novel class of Floquet topological phases. The lattice models with time glide symmetry that we presented can be engineered in periodically driven systems such as cold atoms. Namely, the 2D model consisting of a stack of driven Su-Schrieffer-Heeger model would be realizable by designing superlattice potential that shows alternating hopping amplitude and by employing synthetic gauge fields for imaginary hoppings. The 3D model consisting of a stack of Haldane model on the honeycomb lattice may also become feasible in a near future because the Haldane model has been recently realized in cold atoms by shaking the lattice potential \cite{Jotzu}.

In equilibrium systems, topological crystalline insulators (TCIs) have been studied actively as representative topological materials where gapless surface states are protected by spatial symmetry~\cite{Fu11,Hsieh12,Chiu13,Morimoto13,Shiozaki14}. Since static reflection symmetry also serves as time-glide symmetry in Floquet systems, a natural question would be whether the time-glide symmetric Floquet topological phases are different topological phases from TCIs or Floquet versions of them. In the following, we show that these two are indeed distinct sets of topological phases.
(i) First, the 2D class AIII model for a time-glide symmetric Floquet topological phase presented in Sec.~\ref{subsec:2d AIII} cannot be a TCI.
In order to have nontrivial time glide Floquet phase, we need anticommutation relation of chiral symmetry and time glide ($\{\Gamma, M_T\}=0$). This is in contrast to topological crystalline insulators in 2D class AIII which require commutation relation, $[\Gamma, R]=0$ (since we need to define 1D winding number with $\Gamma$ in each subspace of $R=\pm 1$).
(ii) In general, the classification of time-glide Floquet topological phases in Table~\ref{table: Classification time glide} shows that they are different from TCIs in the cases with particle hole symmetry $C$ (D and C) or with chiral symmetry $\Gamma$ (AIII, BDI, DIII, CII, CI). In these cases, commutation/anticommutation between $C$ and $M_T/R$, or $\Gamma$ and $M_T/R$, are opposite in obtaining nontrivial phases. (For comparison, the classification table for TCIs are found in  Ref.~\cite{Morimoto13} as Table VI.)
(iii) Classes A, AI, AII do not show difference in the classification for time-glide Floquet topological phases and TCIs. However, at least, ones in 3D class A are distinct from TCIs, as discussed in Sec.~\ref{subsec: 3D A}. Namely, the way topological numbers are defined are different. In characterizing time-glide Floquet topological phases, we use half-period unitary $U(0 \to T/2)$ as a function of $k_x$ and $k_y$ and see wrapping of a special point as shown in Fig.~\ref{fig: 3D topological invariant}. On the other hand, characterization of a Floquet version of TCI involves $U(k_x,k_y,t)$ as a function of three parameters which belongs to a nontrivial element of $\pi_3(U(N))=Z$. Since we can consider a case where half-period unitary is nontrivial but full-period unitary is trivial in $\pi_3(U(N))$, time glide Floquet topological phases and TCIs are generally different in 3D class A.
These considerations indicate that time-glide Floquet topological phases and TCIs are distinct sets of topological phases in general.

While we focused on time-glide symmetry $M_T$ and two-fold time-screw symmetry $C_{2T}$ in this paper, we can consider more general time-nonsymmorphic symmetries and Floquet topological phases protected by them.
For example, we can consider time screw symmetry having $C_3$ symmetry which does not square to 1. We expect that such non order-two symmetries also support some nontrivial Floquet topological phases. However, these general nonsymmorphic space-time symmetry cannot be directly incorporated into the Clifford-algebra approach, which is naturally suited of order-two symmetries.
Thus Floquet topological phases protected by general time-nonsymmorphic symmetries are left for future studies. Since twisted equivariant K-theory is proposed to be a framework for classifying equilibrium topological phases with crystalline symmetries~\cite{FreedMoore}, twisted equivariant K-theory may be applicable to Floquet topological phases.

We studied effects of dynamical symmetry on crystals under periodic driving. Recently, it has been proposed in Ref.~\cite{Boyle16} that even crystal structure itself can be extended to a dynamical space-time pattern of atoms, which is called ``choreographic crystals''.
In choreographic crystals, atoms move in a symmetric way with respect to each other within a period, like orbiting satellites. These space-time patterns of atoms were studied by extending group theory to dynamical motions. It would be an interesting future problem to classify these choreographic crystals and also study topological phases realized by electrons sitting on these space-time patterns.

Periodic driving can cause catastrophic heating of the topological phases in a long time when interactions are present in the system.
Therefore, we will primarily be concerned with systems that are well approximated as non-interacting and clean. In fact, even in the presence of relatively weak interactions, heating may only set in at extremely long times \cite{Abanin17}, allowing for a broad pre-thermalization window. 
This heating effect can be entirely avoided when the bulk is many-body localized \cite{Bahri,Chandran,Nandkishore}. 
Although including disorder usually breaks reflection symmetry and it is not so natural to combine time-glide symmetry with disorder, it would still be useful to consider effects of disorder that statistically preserves time-glide symmetry, having in mind that some disordered TCIs that preserves reflection symmetry on average support stable gapless surface states~\cite{Fulga14,Morimoto14}.
It was shown that systems with Abelian symmetry group can be localized by suitable local randomness while those with non-Abelian symmetry cannot be localized \cite{DrewRomain}. In our case, the criterion will be whether the reflection symmetric subsystem can be localized or not when we introduce reflection symmetric random potential. At the reflection symmetric subsystem, time glide $M_T$ reduces to global $Z_2$ symmetry, where we can apply the condition in Ref.~\cite{DrewRomain}. Topological phases in 2D class AIII in Sec.~\ref{subsec:2d AIII} requires $\{\Gamma, M_T\}=0$ and cannot be localized while that in 3D class A in Sec.~\ref{subsec: 3D A} can be localized. In any case, these Floquet topological phases are stable under well-controlled systems when interactions are absent.
Furthermore, along this line, we may consider global $Z_2$ symmetry with half-period time translation as a future problem, because such global symmetry makes it easier to localize the bulk. For example, nontrivial entries in tenfold way with such symmetry commuting with other on-site symmetries host nontrivial Floquet topological phases.  

Finally, we note on effects of interactions onto time glide symmetric Floquet topological phases. Recently, interacting Floquet topological phases have been actively studied \cite{Keyserlingk,ElseNayak,PotterMorimotoVishwanath}. In particular, Floquet topological phases show a breakdown of noninteracting topological phases with interactions which is different from equilibrium cases. This arises from the presence of Floquet evolution operator acting like an effective symmetry operation~\cite{PotterMorimotoVishwanath}. Therefore, it will be interesting to consider such interacting Floquet topological phases with time glide symmetry.


\begin{acknowledgments}
We thank A.~C.~Potter and M.~Serbyn for fruitful discussions.
TM was supported by 
the EPiQS initiative of the Gordon and Betty Moore Foundation. AV was supported by a Simons Investigator grant and AV and HCP were supported by by NSF DMR-1411343.
\end{acknowledgments}

\appendix
\section{Ten-fold way classification of Floquet topological phases \label{app: tenfold way}}
We review tenfold way classification of noninteracting Floquet topological phases obtained in Ref~\cite{Roy16}. We consider topological characterization of a time-evolution operator $U_S(\bm k, t)$. Instead of studying $U_S(\bm k, t)$ itself, we study the effective Hamiltonian $H_S$ in Eq.~(\ref{eq: H_S}) defined with $U_S(\bm k, t)$, because we can apply classification procedure for equilibrium topological phases. 
First we study complex symmetry classes (class A and class AIII).
\begin{enumerate}
\item class A: 
The Hamiltonian $H_S$ obeys the inherent chiral symmetry $\tilde \Gamma$. The Floquet topological phase is mapped to a TI in $(d+1)$D class AIII.

\item class AIII: 
The Hamiltonian $H_S$ obeys the inherent chiral symmetry $\tilde \Gamma$ and a unitary symmetry $\Gamma'$. 
If there is a commuting unitary symmetry, the tenfold way classification should be performed for the Hamiltonian after block diagonalization. 
The block diagonalized Hamiltonian with respect to $\Gamma'=\pm 1$
has no symmetry due to $\{\Gamma', \tilde \Gamma\}$.
The Floquet topological phase is mapped to a TI in $(d+1)$D class A.

\end{enumerate}

Next we study real symmetry classes with either $T$ or $C$ (class AI, class AII, class D and class C).
Since these TRS and PHS acts on $t$ in a trivial way,
the Hamiltonian $H_S(\bm k,t)$ is interpreted as one-parameter family (parameterized by $t$) of $d$-dimensional Hamiltonians characterizing a topological defect \cite{teo-kane10}. 
\begin{enumerate}
\item class AI: 
The Hamiltonian $H_S$ obeys the inherent chiral symmetry $\tilde \Gamma$ and the TRS $T'$ squaring to $+1$. 
Combining these two leads to an effective PHS $\tilde \Gamma T'$ squaring to $-T^2=-1$.
Thus the Floquet topological phase is mapped to a 1D topological defect in a $d$D class CI system which is further mapped to $d-1$D TI in class CI.

\item class AII: 
The Hamiltonian $H_S$ obeys the inherent chiral symmetry $\tilde \Gamma$ and the TRS $T'$ squaring to $-1$. 
Combining these two leads to an effective PHS $\tilde \Gamma T'$ squaring to $-T^2=+1$.
Thus the Floquet topological phase is mapped to a 1D topological defect in a $d$D class DIII system which is further mapped to $d-1$D TI in class DIII.

\item class D: 
The Hamiltonian $H_S$ obeys the inherent chiral symmetry $\tilde \Gamma$ and the TRS $C'$ squaring to $+1$. 
Combining these two leads to an effective PHS $\tilde \Gamma C'$ squaring to $C^2=+1$.
Thus the Floquet topological phase is mapped to a 1D topological defect in a $d$D class BDI system which is further mapped to $d-1$D TI in class BDI.

\item class C: 
The Hamiltonian $H_S$ obeys the inherent chiral symmetry $\tilde \Gamma$ and the TRS $T'$ squaring to $-1$. 
Combining these two leads to an effective PHS $\tilde \Gamma T'$ squaring to $-T^2=+1$.
Thus the Floquet topological phase is mapped to a 1D topological defect in a $d$D class DIII system which is further mapped to $d-1$D TI in class DIII.

\end{enumerate}

Finally we study real symmetry classes with both $T$ and $C$ (class BDI, class DIII, class CII and class CI).
Since the chiral symmetry $\Gamma=TC$ results in a commuting unitary symmetry $\Gamma'$ for $H_S$,
we focus on the block diagonalized Hamiltonian and remaining symmetries to deduce the classification.
The inherent chiral symmetry $\tilde \Gamma$ is no longer a symmetry for the block diagonalized Hamiltonian because $\{\Gamma', \tilde \Gamma \}=0$.
When $\Gamma^2=(TC)^2=+1$ (in class BDI and class CII),
$T'$ and $C'$ remain as symmetries for the block diagonalized Hamiltonian with $\Gamma'=\pm 1$.
Since $T'$ and $C'$ are equivalent after block diagonalization because $T'=\tilde \Gamma C'$,
only one symmetry, say $T'$, remains as a symmetry after block diagonalization.
When $\Gamma^2=(TC)^2=-1$ (in class BDI and class CII),
$T'$ and $C'$ are not symmetries for the block diagonalized Hamiltonian with $\Gamma'=\pm i$ (complex conjugation in $T'$ and $C'$ exchanges two sectors $\Gamma'=\pm i$).
Instead, $\tilde \Gamma T'$ remains as a PHS for for the block diagonalized Hamiltonian with $\Gamma'=\pm 1$.
\begin{enumerate}
\item class BDI: 
The block diagonalized Hamiltonian $H_S$ obeys the TRS $T'$ squaring to $+1$. 
Thus the Floquet topological phase is mapped to a 1D topological defect in a $d$D class AI system which is further mapped to $d-1$D TI in class AI.

\item class DIII: 
The block diagonalized Hamiltonian $H_S$ obeys the PHS $\tilde \Gamma T'$ squaring to $-T^2=+1$. 
Thus the Floquet topological phase is mapped to a 1D topological defect in a $d$D class D system which is further mapped to $d-1$D TI in class D. 

\item class CII: 
The block diagonalized Hamiltonian $H_S$ obeys the TRS $T'$ squaring to $-1$. 
Thus the Floquet topological phase is mapped to a 1D topological defect in a $d$D class AII system which is further mapped to $d-1$D TI in class AII. 

\item class CI: 
The block diagonalized Hamiltonian $H_S$ obeys the PHS $\tilde \Gamma T'$ squaring to $-T^2=-1$. 
Thus the Floquet topological phase is mapped to a 1D topological defect in a $d$D class C system which is further mapped to $d-1$D TI in class C.

\end{enumerate}

To summarize, noninteracting Floquet topological phases in the $d$-dimensional space share the same topological classification as equilibrium topological phases in the same symmetry class.
Corresponding topological number can be defined for each gap in the quasi energy spectrum for symmetry classes without PHS or chiral symmetry; a system with $n$ gaps is characterized by $n$ topological numbers in those classes.
For classes with PHS or chiral symmetry, two topological numbers are defined corresponding to  0 gap and $\pi$ gap. 
The result is summarized in Table~\ref{table: AZ classes}.

\bibliography{floquet}

\end{document}